\documentclass[preprint,eqsecnum]{aastex}
\usepackage{natbib}
\usepackage{graphicx}
\usepackage{amsfonts,amsmath,amssymb}
\usepackage{epstopdf}
\title{Detection, Localization and Characterization of Gravitational Wave Bursts in a Pulsar Timing Array}

\author{Lee Samuel Finn\altaffilmark{1}, Andrea N. Lommen\altaffilmark{2,3}}
\altaffiltext{1}{Department of Physics, Department of Astronomy \& Astrophysics, The Pennsylvania State University, University Park, PA 16802}
\altaffiltext{2}{Franklin \& Marshall College,
Department of Physics and Astronomy, Lancaster, PA 17604}
\altaffiltext{3}{Australia Telescope National Facility, CSIRO, PO BOX 76, Epping NSW 1710, Australia}
\shortauthors{Finn and Lommen}
\subjectheadings{methods: data analysis --- methods: statistical --- gravitational waves}

\begin{abstract}
Efforts to detect gravitational waves by timing an array of pulsars have focused traditionally on stationary gravitational waves: e.g., stochastic or periodic signals.
Gravitational wave bursts --- signals whose duration is much shorter than the observation period --- will also arise in the pulsar timing array waveband. 
Sources that give rise to detectable bursts include the formation or coalescence of supermassive black holes (SMBHs), the periapsis passage of compact objects in highly elliptic or unbound orbits about a SMBH, or cusps on cosmic strings.  
Here we describe how pulsar timing array data may be analyzed to detect and characterize these bursts. 
Our analysis addresses, in a mutually consistent manner, a hierarchy of three questions: \emph{i}) What are the odds that a dataset includes the signal from a gravitational wave burst? \emph{ii}) Assuming the presence of a burst, what is the direction to its source? and \emph{iii}) Assuming the burst propagation direction, what is the burst waveform's time dependence in each of its polarization states? 
Applying our analysis to synthetic data sets we find that we can \emph{detect} gravitational waves even when the radiation is too weak to either localize the source of infer the waveform, and \emph{detect} and \emph{localize} sources even when the radiation amplitude is too weak to permit the waveform to be determined. 
While the context of our discussion is gravitational wave detection via pulsar timing arrays, the analysis itself is directly applicable to gravitational wave detection using either ground or space-based detector data. 
\end{abstract}

\begin{document}

\section{Introduction}


It has been just over thirty years since \citet{sazhin:1978:ofd} and \citet{detweiler:1979:ptm} showed how gravitational waves could be detected by correlating the timing residuals of a collection of pulsars, and twenty years since \citet{foster:1990:cpt} proposed using a collection of pulsars --- i.e., a pulsar timing array --- to achieve greater sensitivity. Over the ensuing years telescope collecting area has increased, antenna temperature has decreased, pulsar timing electronics and methodology has improved, and pulsars with exceptionally low intrinsic timing noise have been discovered. As a result of these advances the near-future detection of a stochastic gravitational wave signal through pulsar timing observations is a strong possibility.  Analyses aimed at detecting gravitational waves using pulsar timing array observations have traditionally focused on stationary signals (i.e., stochastic or periodic gravitational waves). More generally, analyses aimed at detecting gravitational waves have merged the questions of detection and characterization, overlooking the possibility of detecting a signal that is too weak to be characterized. Here we describe how pulsar timing array data may be analyzed to search for gravitational wave \emph{bursts}, demonstrating that \emph{i}) pulsar timing array data is sufficiently rich to allow the detection of gravitational wave bursts, the localization of the burst source, and the time-dependent waveform of the radiation in its (two) polarization states; and, \emph{ii}) that gravitational wave signals too weak to be characterized, or too weak to allow their source to be localized, may still be strong enough to be unambiguously detected. 

The first detections of gravitational waves will be important for confirming their existence and testing whether general relativity correctly predicts their properties (e.g., polarization modes, propagation speed). Of perhaps greater long-term significance will be the use of gravitational waves as a tool of observational astronomy that gives us direct insight into phenomena that we can now observe only indirectly, if at all. For example, \citet{jaffe:2003:gwp,wyithe:2003:lgw} and \citet{jenet:2006:ubo} have shown that root-mean-square (rms) amplitude of a stochastic signal arising from the confusion limit of a large number of supermassive black hole binary coalescences is within an order of magnitude of the current sensitivity of the most advanced pulsar timing array. Since the signals that contribute to this background arise from a population of discrete sources distributed throughout space we quite reasonably expect that some of the individual sources may be observable as gravitational wave bursts rising above this background. Indeed, recent work by \citet{sesana:2008:sgb} shows that at frequencies greater than a few times $10^{-8}$ nanohertz the gravitational ``background'' arising from supermassive binary black hole coalescense should be dominated by a few bright sources. Other potential burst gravitational wave sources in the pulsar timing array band include cosmic (super)string cusps and kinks \citep{damour:2001:gwb, siemens:2007:gsb,leblond:2009:gwf}. 

Pulsar timing array observations are sensitive to gravitational waves of periods ranging from the interval  between timing observations (days to months) and the duration of the observational data sets (years). The corresponding wavelengths are much greater than those explored in existing or proposed human-built ground or space-based detectors. Ground-based detectors, whether of the  acoustic \citep{astone:2010:i1s} or interferometric variety \citep{accadia:2010:sap,riles:2010:pgw}, are currently sensitive to waves in the $\sim$ 100~Hz -- 1~kHz band with proposed advances opening-up the 10--100~Hz band \citep{smith:2009:pte,kuroda:2006:sol}. Space-based detectors, which have been the subject of extensive design studies over the last thirty years, would be sensitive to gravitational waves in the $\sim3\times10^{-5}$~Hz -- 10~Hz band \citep{stebbins:2006:lmt,jennrich:2009:trl,kawamura:2008:jsg}. Over this broad band --- $10^{-9}$ -- $10^3$~Hz --- the scale and character of the sources varies dramatically: e.g., ground-based detectors will be sensitive to gravitational waves from neutron star or solar mass black hole binaries, supernovae and gamma-ray burst progenitors; space-based detectors to gravitational waves from white dwarf binaries, stellar disruptions about intermediate mass black holes and the inspiral of solar mass compact objects or intermediate mass black holes about $10^{4.5}--10^{7.5}\,\mathrm{M}_\odot$ black holes; and pulsar timing arrays to the formation, interaction and evolution of supermassive black holes. Pulsar timing array observations thus offer their own, unique perspective on the gravitational wave universe, distinct from that provided by the either ground- and space-based detectors. 

Analyses aimed at detecting gravitational waves using pulsar timing array observations have traditionally focused on stationary signals: i.e., an isotropic stochastic gravitational wave background \citep{hellings:1983:ulo,mchugh:1996:pta,thorsett:1996:ptl,lommen:2001:pmt,lommen:2003:nlo,jenet:2005:dsg,jenet:2006:ubo,demorest:2007:mgw,hobbs:2008:upt,haasteren:2009:omg,anholm:2009:osf} or gravitational waves from discrete periodic sources \citep{lommen:2001:upt,jenet:2004:cpo,jenet:2005:pta,jenet:2005:cpo}. More recent work \citep{haasteren:2009:gma} has investigated the detection of gravitational wave ``memory'' \citep{christodoulou:1991:nno} associated with sources that radiate a significant amount of energy in gravitational waves \citep{wiseman:1991:cng} or that become unbound \citep{thorne:1992:gbw}.

Gravitational wave detection using a pulsar timing array begins with the collection of timing residuals from the several array pulsars. These timing residuals are the difference between the expected pulse arrival times (taking into account all non-gravitational-\emph{wave} propagation effects) and the actual pulse arrival times at each pulsar observational epoch. For pulsars used in current timing arrays the timing precision is in the 50~ns -- 5~$\mu$s range. In \S\ref{sec:response} we summarize how these timing residuals reflect the passage of a plane gravitational wave through the pulsar-Earth baseline. In \S\ref{sec:methodology} we describe our analysis for gravitational wave bursts, which  takes advantage of the correlation of the timing residuals measured for different pulsars. In \S\ref{sec:examples} we demonstrate the effectiveness of the analysis by applying it to simulated data arising from a thirty pulsar timing array and including a gravitational wave burst such as would be expected from a parabolic encounter of two supermassive black holes. Finally, in \S\ref{sec:conclusions} we summarize our findings and describe planned future work. 

\section{Pulsar timing response to the passage of a gravitational wave burst}
\label{sec:response}
\subsection{Introduction}
A pulsar timing array dataset consists of a collection of pulsar ``time of arrival'', or TOA, measurements for pulses of the individual pulsars that comprise the array. The arrival time observations are made for each pulsar over a period years, with successive pulse arrival time observations for each array pulsar made anywhere from days to months apart. The TOA measurements are compared to predicted arrival times based on timing models for the individual pulsars, which includes all non-gravitational-wave effects that affect the arrival times. The difference between the observed and expected pulse arrival times are referred to as timing residuals, which are then presumed to consist of timing noise and gravitational wave effects. Evidence for gravitational waves is sought in the timing residuals.\footnote{The procedure of fitting the timing model to the pulsar arrival time measurements for gravitational wave analysis has the unfortunate side-effect of ``fitting out'' any gravitational wave contributions that have the form of other timing model effects. We address this point directly in the conclusions.} In this section we calculate the contribution to pulse arrival times owing to a passing plane gravitational wave burst. 

\subsection{Gravitational waves}\label{sec:gravWaves}
Denote the perturbative plane gravitational wave, expressed in transverse-traceless (TT) gauge \citep{misner:1973:g}, as 
\begin{equation}\label{eq:h}
\mathbf{h}(t,\vec{x}) =
{h}_{+}(t-\hat{k}\cdot\vec{x})\mathbf{e}^{(+)}(\hat{k}) +
{h}_{\times}(t-\hat{k}\cdot\vec{x})\mathbf{e}^{(\times)}(\hat{k})
\end{equation}
where $\hat{k}$ is the plane wave propagation direction and $\mathbf{e}^{(+)}$ and $\mathbf{e}^{(\times)}$ are the two independent gravitational wave polarization basis tensors, 
\begin{subequations}
\begin{align}
e_{(+)}^{lm}e^{(+)}_{lm} &= e_{(\times)}^{lm}e^{(\times)}_{lm} = 2\\
e_{(+)}^{lm}\hat{k}_{m} &= e_{(\times)}^{lm}\hat{k}_m = e_{(+)}^{lm}e^{(\times)}_{lm} = 0.
\end{align}
\end{subequations}
Locating the coordinate system origin at the solar system barycentre consider a pulsar at spatial rest located at $\vec{x}_p$, 
\begin{equation}
\vec{x}_p(t) = L\hat{n}
\end{equation}
where $\hat{n}$ is the unit vector in the direction of the pulsar and $L$ is pulsar's distance.

\subsection{Timing residuals}\label{sec:residuals}
Focus attention on the electromagnetic field associated with the pulsed emission of a pulsar and denote the fields phase, at the pulsar, as $\phi_0(t)$. We are interested in the time-dependent phase $\phi(t)$ of the electromagnetic field associated with the pulsed emission measured at an Earth-based radio telescope, which we write as
\begin{subequations}
\begin{equation}
\phi(t) = \phi_0[t-L-\tau_0(t)-\tau_{\mathbf{GW}}(t)]
\end{equation}
where
\begin{align}
\tau_0 &= 
\left(\begin{array}{l}
\text{Corrections owing exclusively to the spatial motion of the Earth}\\
\text{within the solar system, the solar system with respect to the pulsar,}\\
\text{and electromagnetic wave propagation the interstellar medium}
\end{array}\right)\\
\tau_{\mathrm{GW}} &= 
\left(\begin{array}{l}
\text{Corrections owing exclusively to $\mathbf{h}(t,\vec{x})$}
\end{array}\right).
\end{align}
\end{subequations}
(Note that we work in units where $c=G=1$.) 
In the absence of gravitational waves $\tau_{\mathrm{GW}}$ vanishes and the front $\phi_0(t)$ arrives at Earth at time $t_\oplus(t) = t+L+\tau_0(t)$. In the presence of a gravitational wave signal the phase front arrives at time $t_\oplus(t)+\tau_{\mathrm{GW}}(t)$; thus, $\tau_{\mathrm{GW}}$ is the gravitational wave timing residual. Following \citet{finn:2009:roi} Eqs.~(3.26) and (3.12e), the arrival time correction $\tau_{\mathrm{GW}}(t)$ is 
\begin{subequations}\label{eq:tauGW}
\begin{align}
\tau_{\mathrm{GW}}(t)
&= -\frac{1}{2}n^ln^m\left[e^{(+)}_{lm}\mathcal{H}_{(+)}+e^{(\times)}_{lm}\mathcal{H}_{(\times)}\right]
\end{align}
where
\begin{equation}
\mathcal{H}_{(A)}(t,L,\hat{k}_jn^j) = \int_0^L h_{A}\left(t-(1+\hat{k}_jn^j)(L-\lambda)\right)d\lambda
\label{eqn:integral_of_h}
\end{equation}
\end{subequations}
It is convenient to introduce $f_{A}(u)$,
\begin{equation}
\frac{df_{A}}{du} = h_A(u),
\end{equation}
and rewrite equation \ref{eqn:integral_of_h} using $f_{A}(u)$ as follows:
\begin{equation}\label{eq:EarthPsr}
\mathcal{H}_{(A)}(t,L,\hat{k}_jn^j) = \frac{f_{A}(t)}{1+\hat{k}_jn^j} - \frac{f_{A}(t-(1+\hat{k}_jn^j)L)}{1+\hat{k}_jn^j}.
\end{equation}
The contribution proportional to $f_{A}(t)$ is colloquially referred to as the ``Earth'' term; similarly, the contribution proportional $f_{A}\left(t-(1+\hat{k}_m\hat{n}^m)L\right)$ is referred to as the ``Pulsar'' term. The Pulsar term is of central importance when pulsar timing data is used to bound the strength of a stochastic gravitational wave background \citep{jenet:2005:dsg}; however, as we show below, only the Earth term is important when our goal is to use pulsar timing data to detect gravitational wave bursts. 

\subsection{Discussion}

At this point it is worth noting several properties of the timing residual $\tau_{\mathrm{GW}}$. 

\subsubsection{Burst detection involves only the Earth term}\label{sec:et}
As shown in equation (\ref{eq:EarthPsr}) the gravitational wave induced timing residuals for any pulsar may be written as the difference of two functions,  each of which is an \emph{integral} of $h_{+,\times}(t,\vec{x})$. These two functions are identical, except that one is displaced in time by an amount $L(1+\hat{k}_m\hat{n}^m)$ with respect to the other. Correspondingly, 
\begin{itemize}
\item When timing residual measurements from an array of pulsars are available the first evidence for the passage of a gravitational wave burst will appear simultaneously in all observed residuals; and  
\item As long as the burst duration $\Delta T$ and the observation duration $T$ are less than $(1+\hat{k}_m\hat{n}^m)L/c$ only the Earth term contributes to the \emph{correlated} timing residuals in the pulsar timing array.\footnote{Other bursts, having interacted with individual pulsars at much earlier times (thousands of years) will contribute to the timing noise of individual pulsars. These contributions will \emph{not} be correlated among the pulsars in the timing array over the human observational timescale (decades).} 
\end{itemize}
When searching for gravitational wave bursts we can thus ignore the pulsar term except for sources within an angle 
\begin{align}
\theta_p &< \cos^{-1}\left[1-\frac{\Delta T}{L_p}\right] \sim 2.5\,\deg \left[\frac{\Delta T}{1\,\mathrm{yr}}\frac{1\,\mathrm{kpc}}{L}\right]^{1/2}
\end{align}
of pulsar $p$. 

\subsubsection{Timing residuals in a pulsar network are sensitive to gravitational wave polarization}\label{sec:polSens}
The timing residual $\tau_{\mathrm{GW}}$ is a linear combination of the integrals of the two polarizations of the waveform  $\mathcal{H}_{(A)}$: i.e., we may rewrite \ref{eq:tauGW} as
\begin{equation}
\tau_{\mathrm{GW}}(t) =-\frac{1}{2}\left[F^+\mathcal{H}_{(+)}+F^{\times}\mathcal{H}_{(\times)}\right],
\end{equation}
where
\begin{equation}
F^{(A)} = \hat{n}^l\hat{n}^m\mathbf{e}^{(A)}_{lm}(\hat{k}).
\end{equation}
The timing residual correlations of PTA pulsars take a form that depends on the pulsar locations and the gravitational wave polarization. When the wave propagation direction $\hat{k}$ is known the measured timing residuals of two appropriately chosen pulsars is sufficient to measure separately the radiation in each of the two gravitational wave polarization states.

\subsubsection{Timing residuals in a pulsar timing array are sensitive to wave propagation direction.}\label{sec:kSens}
The polarization tensors $\mathbf{e}^{(A)}_{lm}$ are orthogonal to the wave propagation direction $\hat{k}$; correspondingly, the relative contribution of the $\mathcal{H}_{(A)}$ to the timing residual $\tau_{\mathrm{GW}}$ for a given pulsar depends on the gravitational wave propagation direction through the  $F^{(A)}$. In addition, the overall amplitude of the timing residual for any particular pulsar depends on the wave propagation direction through the additional factor $(1+\hat{k}_m\hat{n}_p^m)^{-1}$. Observation of the timing residuals in three pulsars, with appropriately chosen lines-of-sight from Earth, are thus sufficient to measure the radiation propagation direction. 

Combining the insights of subsections \ref{sec:et}, \ref{sec:polSens} and \ref{sec:kSens} we see that a pulsar timing array of five or more pulsars has, in principle, sufficient information to fully characterize a passing gravitational wave burst. In the following section we describe the statistical methodology by which we can infer $\hat{k}$ and $h_{+,\times}(t)$ at, e.g., the solar system barycentre from the measured timing residuals in a pulsar timing array. 

\subsubsection{Pulsar timing residuals are larger for longer bursts than for shorter bursts}
The gravitational wave induced timing residual associated with any particular pulsar is proportional to the \emph{integral} of $h_{ij}(t)$ over time (see eq. \ref{eqn:integral_of_h}). This leads to an important point: for fixed strain amplitude and waveform ``shape'', the timing residuals associated with bursts have magnitudes proportional to the burst duration. This is very different than is the case with ground-based gravitational wave detectors (e.g., the Laser Interferometer Gravitational Wave Observatory (LIGO) \citep{saulson:1994:foi}) or the proposed space-based detector LISA, where the measured quantity responds directly to the gravitational wave strain. The difference arises because the gravitational wave signal band of interest for ground- and space-based detectors has wavelengths greater than the detector size, while the band of interest for pulsar timing array measurements has wavelengths much smaller than the detector size (i.e., the pulsar-Earth baseline distance).\footnote{For LISA the detector bandwidth does extend to wave frequencies a few times greater than the round-trip travel along the $5\times10^6$~km arm baseline. This effect of greater sensitivity at longer periods is apparent in the high-frequency part of LISA's response function.}

\section{Statistical Methodology}
\label{sec:methodology}

\subsection{Framing the questions}

Our goal is three-fold. First, ascertain the odds that the particular data set $\mathbf{d}$ includes a contribution characteristic of a passing gravitational wave burst; second, assuming that is so, determine the probability that the contribution is characteristic of a wave propagating in the direction $\hat{k}$; and, finally, assuming the contribution is characteristic of a burst propagating in direction $\hat{k}$, determine the probability that the contribution is characteristic of a waveform at Earth described by $\mathbf{h}= h_{+}(t-\hat{k}\cdot\vec{x})\mathrm{e}_{+}(\hat{k})  + h_{\times}(t-\hat{k}\cdot\vec{x})\mathrm{e}_{\times}(\hat{k})$ for functions $h_{+}$ and $h_{\times}$. 

While actual analysis might address these questions in the order given above it is pedagogically simpler and more instructive to approach them in the opposite order, which we do in the three subsections that follow.

\subsection{Inferring $\mathbf{h}$}\label{sec:infer}

Given timing residual observations $\mathbf{d}$ from an array of pulsars that include a contribution from a plane gravitational wave propagating past Earth in direction $\hat{k}$, what is the probability $p_h$ that the wave is described by $\mathbf{h}$? 

The desired probability density depends on the response of the pulsar network to incident gravitational waves, the statistical properties of the measurement and intrinsic pulsar timing noise, and the assumed direction of wave propagation:
\begin{subequations}
\begin{align}
p_{h}(\mathbf{h}|\hat{k},\mathcal{I}) &= 
\left(\begin{array}{l}
\text{probability that gravitational wave burst is described by the wave $\mathbf{h}$}\\
\text{propagating in direction $\hat{k}$, and other, unenumerated assumptions $\mathcal{I}$}
\end{array}\right).
\end{align}
\end{subequations}
Exploiting the Bayes' Theorem, the probability density $p_{h}$ can be expressed in terms of the normalized likelihood $\Lambda$, an a priori probability density $q_h$ that expresses expectations regarding $\mathbf{h}$, and a normalization constant $Z_h$, often referred to as the evidence:
\begin{subequations}\label{eq:ph}
\begin{align}
p_{h}(\mathbf{h}|\mathbf{d},\hat{k},\mathcal{I}) &= 
\frac{\Lambda(\mathbf{d}|\mathbf{h},\hat{k},\mathcal{I})
q_{h}(\mathbf{h}|\hat{k},\mathcal{I})}{Z_h(\mathbf{d}|\hat{k},\mathcal{I})}\\
\Lambda(\mathbf{d}|\mathbf{h},\hat{k},\mathcal{I}) &=
\left(\begin{array}{l}
\text{Probability of observing TOA residuals $\mathbf{d}$ given}\\
\text{gravitational wave $\mathbf{h}$ propagating in direction $\hat{k}$}
\end{array}\right)
\\
q_{h}(\mathbf{h}|\hat{k},\mathcal{I})&=
\left(\begin{array}{l}
\text{a priori probability density that $\mathbf{h}$ describes the}\\
\text{gravitational wave burst propagating in direction $\hat{k}$}
\end{array}\right)
\\
Z_h(\mathbf{d}|\hat{k},\mathcal{I})&= \int d^n h_{+}\,d^{n}h_{\times}\,\, \Lambda(\mathbf{d}|\mathbf{h},\hat{k},\mathcal{I})q_{h}(\mathbf{h}|\hat{k},\mathcal{I})\nonumber\\
&= p_d(\mathbf{d}|\hat{k},\mathcal{I}) \label{eq:Zh}
\\
&=
\left(\begin{array}{l}
\text{Probability of observing $\mathbf{d}$ assuming the presence of}\\
\text{gravitational wave burst $\mathbf{h}$ propagating in direction $\hat{k}$}
\end{array}\right)\nonumber
\end{align}
\end{subequations}
(In equation \ref{eq:Zh} the integral is over all possible values of the waveform $h_{+}$ and $h_{\times}$ at the $n$ sample times.)
We discuss each of these terms in more detail below. 

\subsubsection{The Likelihood $\Lambda$}\label{sec:likelihood}
Focus attention on pulsar $j$, whose measured timing residuals are represented as the time-series vector $\mathrm{d}_j$. These residuals are the sum of 
of measurement noise, intrinsic pulsar timing noise, scintillation and other propagation noises $n_j$ and the pulse arrival time disturbance owing to the passing gravitational wave. Denoting by $\mathrm{R}_j$ the the timing residual response function for pulsar $j$ the net timing residual measured for pulsar $j$ is
\begin{equation}
\mathrm{d}_j = \mathrm{n}_j + \mathrm{R}_j\mathbf{h}.
\end{equation}
The noise associated with individual pulsar timing residual observations are generally well-modeled as Gaussian distributed with zero mean; correspondingly, the noise associated with collection of observations $\mathrm{d}_j$ are described by a zero-mean multi-variate Gaussian. Denoting the noise auto-correlation for pulsar $j$ as $c_{j}(t_l-t_m)$ write the probability density of observing residuals $\mathrm{d}_{j}$ in the timing data of pulsar $j$ as
\begin{subequations}\label{eq:13}
\begin{align}
\Lambda_j(\mathrm{d}_j|\mathbf{h},\hat{k},\mathcal{I}) &= N(\mathrm{d}_j-\mathrm{R}_j\mathbf{h}|\mathrm{C}_j),
\end{align}
where $\mathrm{C}_j$ is the noise auto-corrleation in detector $j$ and
\begin{align}
N(\mathrm{x}|\mathrm{C}) &=
\left(\begin{array}{l}
\text{(multivariate) normal distribution for zero}\\
\text{mean random deviate $\mathrm{x}$ given co-variance $\mathrm{C}$}
\end{array}\right)\\
&= \frac{
\exp\left[
-\frac{1}{2}
\mathrm{x}^T\mathrm{C}^{-1}\mathrm{x}\right]}{
\sqrt{\left(2\pi\right)^{\dim\mathrm{x}}\det||\mathrm{C}||}}.
\end{align}
\end{subequations}
Recall that the noise covariance $C$ has elements
\begin{align}
C_{jk} = <n(t_j)n(t_k)>
\end{align}
where $n(t)$ is the noise at time $t$ and $<>$ denotes an ensemble average over the noise. Expressed as a function of $\tau=t_k-t_j$, $C(\tau)$ is the noise auto-correlation function, which is just the cosine-transform of, and thus entirely equivalent to, the noise power spectral density \citep{kittel:1958:esp}. White, pink, red, or more complex noise timing noise spectra are thus equally well described by Equation (\ref{eq:13}). 

Now assume that the timing noise associated with the observations $\mathbf{d}_j$ of the $n_p$ different pulsars are uncorrelated. Under this assumption the probability density of a set of timing residuals $\mathbf{d}$, consisting of residuals $\mathrm{d}_j$ from each pulsar $j$ in the network, is 
\begin{subequations}\label{eq:14}
\begin{align}
\Lambda(\mathbf{d}|\mathbf{h},\hat{k},\mathcal{I})
=& \prod_{j=1}^{n_p}\Lambda_j(\mathrm{d}_j|\mathbf{h},\hat{k},\mathcal{I})\\
=& N(\mathbf{d}-\mathbf{R}\mathbf{h}|\mathbf{C}). 
\end{align}
\end{subequations}

\subsubsection{The prior $q_h$}
The a priori probability density $q_h$ describes our expectations, before interpreting the observations $\mathbf{d}$, regarding the gravitational wave burst $\mathbf{h}$. It is often the case that discussions of priors like these are more heated and intense than is warranted by the difference any reasonable choice makes to the final result. To understand how this is so it is worthwhile to return for a moment to Equation (\ref{eq:ph}). The probability $p_h$ is the product of two $h$-dependent terms, $\Lambda$ and $q_h$. All of the data dependence is encapsulated in the likelihood $\Lambda$; i.e., the prior $q_h$ is independent of the observations $\mathbf{d}$. When the data are conclusive $\Lambda$ is more sharply peaked than $q_h$ and the dependence of $p_h$ on $\mathbf{h}$ is dominated by the data-dependent term $\Lambda$. In this case the prior $q_h$ is approximately constant over the volume of $h$ where $p_h$ is large and the particular choice of prior is unimportant. On the other hand, when the data are inconclusive the dependence of $p_h$ on $\mathbf{h}$ is dominated by the prior $q_h$ and the structure of $\Lambda$ is unimportant. \emph{As long as the prior, viewed by itself, does not reflect an overly strong set of expectations about $\mathbf{h}$ it will not matter what particular form it takes except at the margins where the observations are suggestive but not conclusive.} With this in mind we consider the basic assumptions we make regarding a gravitational wave burst and how those are represented in $q_h$. 

To begin, we make no assumption that  the nature of the burst should be correlated with its direction of propagation; i.e., we drop the dependence of $q_h$ on $\hat{k}$:
\begin{equation}
q_{h}(\mathbf{h}|\hat{k},\mathcal{I}) = q_{h}(\mathbf{h}|\mathcal{I}) .
\end{equation}
We also assume that there is no a priori correlation between the two dynamically independent polarization states, in which case
\begin{equation}
q_h(\mathbf{h}|\mathcal{I}) = q_{+}(h_{+}|\mathcal{I}) q_{\times}(h_{\times}|\mathcal{I})
\end{equation}
where the $+$ and $\times$ subscripts denote any two orthogonal polarization states. 
Since the resolution of a gravitational wave into orthogonal polarization states is determined only up to a rotation about the propagation direction it must be the case that $q_+$ and $q_\times$ are the same function $q_0$ of their arguments: i.e., 
\begin{equation}
q_h(\mathbf{h}|\mathcal{I}) = q_{+}(h_{+}|\mathcal{I}) q_{\times}(h_{\times}|\mathcal{I}) = 
q_0(h_{+}|\mathcal{I})
q_0(h_{\times}|\mathcal{I})
\end{equation}
Now suppose we represent the gravitational waveform $\mathbf{h}$ by the values of $h_{+}$ and $h_{\times}$ at the solar system barycentre sampled at $n_h$ times $t_j$:
\begin{subequations}
\begin{align}
h_{+,j} &= h_{+}(t_j)\\
h_{\times,j} &= h_{\times}(t_j). 
\end{align}
\end{subequations}
Assuming that the product $h_{+}(t)h_{+}(t+\tau)$ (similarly $h_{\times}(t)h_{\times}(t+\tau)$) vanishes for $\tau\neq0$ when averaged over the ensemble of all possible waveforms $h_{+}$ ($h_{\times})$, \citet{summerscales:2008:mef} showed that we obtain a functional equation for $q_0$ whose solution is 
\begin{subequations}\label{eq:q0}
\begin{align}
q_0(\vec{h}|\sigma,\mathcal{I}) &= N(\vec{h}|\sigma\mathrm{I})\\
& = 
\left[\left(2\pi\sigma^2\right)^{\dim h}\right]^{-1/2}
\exp\left(-\frac{1}{2}\sum_{k=1}^{\dim h}\frac{h_{k}^2}{\sigma^2}\right)
\end{align}
\end{subequations}
where $\sigma$ is an undetermined constant and $\mathrm{I}$ denotes the appropriately dimensioned identity matrix. . 

Our minimal assumptions thus fix the prior $q_h$ up to two undetermined constants $\sigma_{+}$ and $\sigma_{\times}$:
\begin{align}
q_h(\mathbf{h}|\sigma_{+},\sigma_{\times},\mathcal{I}) &=
N(\vec{h}_{+}|\sigma_{+}\mathrm{I})
N(\vec{h}_{\times}|\sigma_{\times}\mathrm{I}). 
\end{align}
In the statistics literature the new constants $\sigma_{+}$ and $\sigma_{\times}$ are referred to as hyperparameters \citep[Chapter 5]{gelman:2004:bda}. Often times the hyperparameters may have a physical interpretation that allows their values to be set, or a priori probability distributions (hyperpriors) selected to describe them, in which case the hyperparameters are treated on par with the other problem parameters. In our case there is a natural interpretation of $\sigma_{+,\times}$ as the root-mean-square amplitude of the gravitational wave burst. This interpretation is not sufficient to determine $\sigma_{+,\times}$ a priori or determine an a priori probability density over the $\sigma_{+,\times}$. This situation is not at all uncommon. Several methods have been suggested and investigated for the treatment of hyperparameters in this case \citep{galatsanos:1992:mfc,thompson:1993:osb,keren:1996:dpf,mackay:1996:hoo,galatsanos:1998:heu,mackay:1999:coa,cawley:2007:pod}. Comparative studies suggest that the best treatment assigns to the hyperparameters those values that optimize the evidence $Z_h$ regarded as a function of the hyperparameters \citep{mackay:1996:hoo,mackay:1999:coa,molina:1999:bar}. We adopt this procedure here. 

The evidence $Z_h$ is the integral of $\Lambda q_h$ over all $h_{+}$, $h_{\times}$ (see Eq.\ \ref{eq:Zh}). Since all of the probability densities that arise in our problem are normal distributions the evidence may be computed in closed form. Combining Eq.~\ref{eq:Zh} with Eqs.~\ref{eq:13}, \ref{eq:14} and \ref{eq:q0} and completing the square in the exponential we obtain:
\begin{subequations}\label{eq:Zha}
\begin{align}
Z_{h}(\mathbf{d}|\sigma_{+},\sigma_{\times},\mathcal{I}) &=
\frac{\exp\left[-\frac{1}{2}\mathbf{d}^T\mathbf{C}^{-1}\mathbf{d}\right]}{\sqrt{(2\pi)^{\dim \mathbf{d}}{\det||\mathbf{C}||}}}
\frac{
\exp\left[
\frac{1}{2}\left(\mathbf{R}^T\mathbf{C}^{-1}\mathbf{d}\right)^TA^{-1}\left(\mathbf{R}^T\mathbf{C}^{-1}\mathbf{d}\right)
\right]
}{
\sqrt{\det||\mathbf{A}||\sigma^{2\dim h_{+}}_{+}\sigma^{2\dim h_{\times}}_{\times}}
}
\end{align}
where 
\begin{align}
\mathbf{A} &= 
\left(\begin{array}{cc}
\sigma_{+}^{-2} \mathrm{I}_{+}&0\\
0&\sigma_{\times}^{-2}\mathrm{I}_{\times}
\end{array}\right) 
+
\mathbf{R}^T\mathbf{C}^{-1}\mathbf{R}\label{eq:A}
\end{align}
\end{subequations}
and $\mathrm{I}_{+,\times}$ represent the appropriately dimensioned unity matrices.

\subsubsection{The probability density $p_h$}
To summarize, the posterior probability density $p_h$ is given by 
\begin{subequations}\label{eq:pha}
\begin{align}\label{eq:pha0}
p_h(\mathbf{h}|\hat{k},\mathbf{d},\sigma_{+},\sigma_{\times},\mathcal{I}) &=
\sqrt{\frac{\det||\mathbf{A}||}{\left(2\pi\right)^{\dim\mathbf{h}}}}
\exp\left[-\frac{1}{2}\left(\mathbf{h}-\mathbf{h}_0\right)^T\mathbf{A}\left(\mathbf{h}-\mathbf{h}_0\right)\right]
\end{align}
where $\mathbf{h}_0$ satisfies
\begin{align}\label{eq:h0}
\mathbf{A}\mathbf{h}_0 &= \mathbf{R}^T\mathbf{C}^{-1}\mathbf{d}
\end{align}
\end{subequations}
with $\mathbf{A}$ given by Equation \ref{eq:A}. 

The reader may note that Equation \ref{eq:h0} for $\mathbf{h}_0$ bears a superficial resemblance to a ``(regularized) least squares'' estimate for the incident wave. This resemblance is an accident of the notation. The operator $\mathbf{A}$ that appears in Equation \ref{eq:h0} would be a constant in a least squares, or regularized least squares, analysis. Here, however, the regularization constants $\sigma_{+}$ and $\sigma_{\times}$ that appear in $\mathbf{A}$ get their values through the optimization of the evidence $Z_h$, which involves both $\mathbf{A}$ and the observations $\mathbf{d}$. Equation \ref{eq:h0} for $\mathbf{h}_0$ must be solved simultaneously with the optimization of $Z_h$, leading to $\sigma_{+}$, $\sigma_{\times}$ and $\mathbf{h}_0$ that differ from any ``least squares'' analysis. Finally, the principal result of our analysis --- i.e., Equation \ref{eq:pha0} for $p_h$ --- would \emph{never} arise from a least squares (or a maximum likelihood) analysis. 

As is apparent from Eq.\ \ref{eq:pha}, $\mathbf{h}_0$ is \emph{the} waveform that maximizes the probability density $p_h$. As such it is naturally the ``best guess'' for $\mathbf{h}$. The availability of the overall probability density $p_h$ gives us the opportunity to say and do much more. With $p_h$ comes the ability to characterize the certainty should assign to this inferrence and, in general, the ability to propagate errors through any inferences that depend on our estimate for $\mathbf{h}$. (See, for example, \citet{bondarescu:2010:egw}, where $p_h$ is used to estimate the uncertainty in the gravitational wave Stokes Parameters.)

\subsection{Inferring the wave propagation direction $\hat{k}$}\label{sec:localize}

Given timing residual observations $\mathbf{d}$ from an array of pulsars that is assumed to include the signal from a plane gravitational wave propagating past Earth in an unknown direction, what is the probability that the wave is propagating in direction $\hat{k}$?

The desired probability depends on the response of the pulsar network to incident waves and the statistical properties of the measurement and intrinsic timing noise:
\begin{equation}
p_k(\hat{k}|\mathbf{d},\mathcal{I}) = 
\left(\begin{array}{l}
\text{probability that burst is propagating}\\
\text{in the direction $\hat{k}$, given data $\mathbf{d}$ and}\\
\text{other, unenumerated assumptions $\mathcal{I}$}
\end{array}\right)
\end{equation}
Exploiting Bayes' Theorem the probability density $p_k$ can be expressed in terms of $p_{d}$, an a priori probability density that expresses our assumptions regarding $\hat{k}$, and a new normalization constant, referred to as the evidence for $\hat{k}$:
\begin{subequations}
\begin{align}
p_k(\hat{k}|\mathbf{d},\mathcal{I}) 
&= Z^{-1}_k(\mathbf{d}|\mathcal{I})p_d(\mathbf{d}|\hat{k},\mathcal{I}) q_k(\hat{k}|\mathcal{I})
\end{align}
where
\begin{align}
q_k(\hat{k}|\mathcal{I}) &= 
\left(\begin{array}{l}
\text{a priori probability that the gravitational}\\
\text{wave burst is propagating in direction $\hat{k}$}
\end{array}\right)\\
Z^{-1}_k(\mathbf{d}|\mathcal{I}) 
&= \int d^2\Omega_k\,p_d(\mathbf{d}|\hat{k},\mathcal{I}) q_k(\hat{k}|\mathcal{I})
\end{align}
Noting that 
\begin{align}
p_d(\mathbf{d}|\hat{k},\mathcal{I})
&= \int d^nh_{+}d^n\,h_{\times}\,\, p_{dh}(\mathbf{d},\mathbf{h}|\hat{k},\mathcal{I})\\
&= \int d^nh_{+}d^n\,h_{\times}\,\, p_h(\mathbf{h}|\mathbf{d},\hat{k},\mathcal{I})Z_h(\mathbf{d}|\hat{k},\mathcal{I})\\
&= Z_h(\mathbf{d}|\hat{k},\mathcal{I})
\end{align}
we find
\begin{align}
p_k(\hat{k}|\mathbf{d},\mathcal{I}) 
&= \frac{
Z_h(\mathbf{d}|\hat{k},\mathcal{I})
}{
Z_k(\mathbf{d}|\mathcal{I})
}
q_k(\hat{k}|\mathcal{I})
\end{align}
where
\begin{align}
Z_k(\mathbf{d}|\mathcal{I}) 
&= \int d^2\Omega_k\,Z_h(\mathbf{d}|\hat{k},\mathcal{I}) q_k(\hat{k}|\mathcal{I})
\end{align}


\end{subequations}
A non-controversial choice of prior $q_k$ arises from assuming that we have no a priori reason to believe that gravitational wave bursts are propagating in any direction preferentially, in which case $q_k$ is uniform on the sphere (i.e., $q_k(\hat{k})=(4\pi)^{-1}$). In that case 
$q(\hat{k}|\mathcal{I})$ is independent of $\mathbf{h}$ and we have
\begin{subequations}\label{eq:pka}
\begin{align}
p_k(\hat{k}|\mathbf{d},\mathcal{I}) &= 
\frac{1}{4\pi}
\frac{
Z_h(\mathbf{d}|\hat{k},\mathcal{I})
}{
Z_k(\mathbf{d}|\mathcal{I})
}\\
Z_k(\mathbf{d}|\mathcal{I}) &= \frac{1}{4\pi}\int d^2\Omega_k\,Z_h(\mathbf{d}|\hat{k},\mathcal{I})\label{eq:Zk}
\end{align}
\end{subequations}
with $Z_h$ given by equation (\ref{eq:Zh}). 

\subsection{Inferring the odds that a gravitational wave is present}\label{sec:detect}
\subsubsection{Model comparison and the Bayes Factor}\label{sec:bayesFactor}

Given timing residual observations $\mathbf{d}$ from an array of pulsars, what odds should we give that a plane gravitational wave was incident on Earth over the period of the observation? 

We treat this question as a problem in Bayesian model comparison \citep{mackay:1992:bi,clark:2007:esm}. The models being compared are a single gravitational wave signal present, denoted $\mathrm{M}_1$, and no gravitational wave signals present, denoted $\mathrm{M}_0$.\footnote{Note that two or more signals present, or noise character changes, or $\ldots$, are all different hypotheses.} Introduce the odds-ratio $\mathcal{O}$ as the the ratio of the probability of hypothesis $\mathrm{M}_1$ to the probability of the hypothesis $\mathrm{M}_0$:
\begin{subequations}
\begin{equation}
\mathcal{O} = \frac{
p_{\mathrm{M}}(\mathrm{M}_1|\mathbf{d},\mathcal{I})
}{
p_{\mathrm{M}}(\mathrm{M}_0|\mathbf{d},\mathcal{I})
}
\end{equation}
where
\begin{align}
p_{\mathrm{M}}(\mathrm{M}_k|\mathbf{d},\mathcal{I}) &= 
\left(\begin{array}{l}
\text{probability, given observations}\\
\text{$\mathbf{d}$, that hypothesis $\mathrm{M}_k$ is true}
\end{array}\right)
\end{align}
\end{subequations}
and $\mathcal{I}$ denotes additional, unenumerated conditions.
Following Bayes' Theorem each of these probabilities can be expressed in terms of a likelihood and an appropriate a priori probability:
\begin{subequations}\label{eq:pM}
\begin{align}
p_{\mathrm{M}}(\mathrm{M}_1|d,\mathcal{I}) &= 
\frac{
q_{\mathrm{M}}(\mathrm{M}_1|\mathcal{I})
}{
Z_{\mathrm{M}}(\mathbf{d}|\mathcal{I})
}
\int d^n\theta\,\Lambda(\mathbf{d}|\mathrm{M}_1,\boldsymbol{\theta},\mathcal{I})q_{\boldsymbol{\theta}\mathrm{M}}(\boldsymbol{\theta}|\mathrm{M}_1,\mathcal{I})\\
p_{\mathrm{M}}(\mathrm{M}_0|d,\mathcal{I}) &= 
\frac{q_{\mathrm{M}}(\mathrm{M}_0|\mathcal{I})}{Z_M(\mathbf{d}|\mathcal{I})}
\Lambda(\mathbf{d}|\mathrm{M}_0,\mathcal{I})
\end{align}
where 
\begin{align}
\Lambda(\mathbf{d}|\mathrm{M}_1,\boldsymbol{\theta},\mathcal{I}) &= 
\left(\begin{array}{l}
\text{probability of observing $\mathbf{d}$ assuming the gravitational}\\
\text{wave signal described by the parameters $\boldsymbol{\theta}$ is present}
\end{array}
\right )\\
\Lambda(\mathbf{d}|\mathrm{M}_0,\mathcal{I}) &= 
\left(\begin{array}{l}
\text{probability of observing $\mathbf{d}$ assuming no signal is present}
\end{array}
\right )\\
q_{M}(\mathrm{M}_k|\mathcal{I}) &= 
\left(\begin{array}{l}
\text{a priori probability of hypothesis $\mathrm{M}_k$}
\end{array}
\right )\\
q_{\theta\mathrm{M}}(\boldsymbol{\theta}|\mathrm{M}_k,\mathcal{I}) &=\left(\begin{array}{l}
\text{a priori probability that $\mathbf{h}$ is described}\\
\text{by parameters $\boldsymbol{\theta}$ given hypothesis $\mathrm{M}_k$}
\text{}
\end{array}
\right )\\
Z_M(\mathbf{d}|\mathcal{I}) &= \sum_k p_{\mathrm{M}}(\mathrm{M}_k|\mathbf{d},\mathcal{I})
\end{align}
\end{subequations}
The odds-ratio $\mathcal{O}$ can thus be expressed as the product of two terms, one that depends only on the observations and one that depends only on our a priori assumptions about the outcome:
\begin{subequations}
\begin{equation}
\mathcal{O} = B(\mathbf{d})\rho
\end{equation}
where 
\begin{align}
B(\mathbf{d}) &= \frac{\int d^n\theta\,\Lambda(\mathbf{d}|\mathrm{M}_1,\boldsymbol{\theta},\mathcal{I})q_{\boldsymbol{\theta}\mathrm{M}}(\boldsymbol{\theta}|\mathrm{M}_1,\mathcal{I})}{\Lambda(\mathbf{d}|\mathrm{M}_0,\mathcal{I})}
\label{eq:B(d)}\\
\rho &= \frac{q_{\mathrm{M}}(\mathrm{M}_1|\mathcal{I})}{q_{\mathrm{M}}(\mathrm{M}_0|\mathcal{I})}
\end{align}
\end{subequations}

$B(\mathbf{d})$, the data-dependent contribution to $\mathcal{O}$, is referred to as the Bayes Factor \citep[pp.\ 184--186]{gelman:2004:bda}. The Bayes Factor reflects the evidence provided by the data $\mathbf{d}$ in favor of the hypothesis $\mathrm{M}_1$ relative to $\mathrm{M}_0$. It is the ratio of the marginalized likelihood of the data under the two hypotheses $\mathrm{M}_1$ and $\mathrm{M}_0$. When it is large compared  to unity the observations favor $\mathrm{M}_1$; when it is small compared to unity the observations favors $\mathrm{M}_0$. The ``odds'' $\mathcal{O}$ are the produce of the Bayes factor and the priors $\rho$. Depending on our interest or prejudice $\rho$ can take on different values. For example, if our interest is to ``let the data speak for themselves'' then we take $\rho=1$; i.e., we expresses no prejudice regarding the presence of absence of a gravitational wave burst in the data set $\mathbf{d}$. Alternatively, if our interest is to express the odds in the context of some theoretical model or prejudice suggesting a rate of gravitational wave bursts over the period of the observation, then $\rho$ will be a function of the rate and observation period. In any case, however, $B$ should be much greater than $\rho^{-1}$ before we are entitled to conclude with certainty that we have observed a gravitational wave burst. 

\subsubsection{Computing the Bayes Factor}
Turn now to computing the Bayes Factor $B(\mathbf{d})$ (Eq.\ \ref{eq:B(d)}). Focus first on the denominator $\Lambda(\mathbf{d}|\mathrm{M}_0)$; i.e., the probability that the particular observation $\mathbf{d}$ is an instance of detector network noise. Referring to the discussion of \S\ref{sec:likelihood} this probability density is
\begin{subequations}
\begin{align}
\Lambda(\mathbf{d}|\mathrm{M}_0,\mathcal{I}) &= 
N(\mathbf{d}|\mathbf{C})\\
&= \frac{
\exp\left[-\frac{1}{2}\mathbf{d}^T\mathbf{C}^{-1}\mathbf{d}\right]
}{
\sqrt{\left(2\pi\right)^{\dim\mathbf{d}}\det||\mathbf{C}||}
}.
\end{align}
\end{subequations}

Turn now to the Bayes Factor numerator,
\begin{equation}\label{eq:marginalization}
\int d^n\theta\,\Lambda(\mathbf{d}|\mathrm{M}_1,\boldsymbol{\theta},\mathcal{I})q_{\boldsymbol{\theta}\mathrm{M}}(\boldsymbol{\theta}|\mathrm{M}_1,\mathcal{I}),
\end{equation}
which we recognize, upon inspection, as the evidence $Z_k(\mathbf{d}|\mathcal{I})$ defined in equation (\ref{eq:Zk}).

The Bayes Factor is thus given by 
\begin{subequations}\label{eq:B(d)a}
\begin{align}
B(\mathbf{d}) &= \frac{\int Z_k(\mathbf{d}|\mathcal{I})}{\Lambda(\mathbf{d}|\mathrm{M}_0,\mathcal{I})}\\
&= \int \frac{d^2\Omega_k}{4\pi}
\frac{\exp\left\{
\frac{1}{2}\left(\mathbf{R}^T\mathbf{C}^{-1}\vec{d}\right)^TA^{-1}\left(\mathbf{R}^T\mathbf{C}^{-1}\vec{d}\right)
\right\}}{\sqrt{\det||\mathbf{A}||\sigma_{+}^{2\dim{h}_{+}}\sigma_{\times}^{2\dim{h}_{\times}}}}
\end{align}
\end{subequations}
where we have taken advantage of the expression for $Z_h$ given in equation (\ref{eq:Zha}). 

\subsection{Summary}
In the preceding discussion we have described a Bayesian analysis that addresses three questions: 
\begin{enumerate}
\item Does the data set $\mathbf{d}$ include the signal from a passing gravitational wave burst?
\item Assuming that a gravitational wave burst is present, what is the probability that the wave is propagating in the direction $\hat{k}$?
\item Assuming a burst propagating in direction $\hat{k}$, what is the probability that the wave at Earth is characterized by  $\mathbf{h}$?
\end{enumerate}
The answers to these questions --- i.e., the principal results of this section --- are given by, for the first question, Equation \ref{eq:B(d)a}; for the second question, Equation \ref{eq:pka}; and, for the third question, Equation \ref{eq:pha}. In the next Section we will demonstrate the effectiveness of this analysis, making use of these three results. 

\section{Examples}
\label{sec:examples}

\subsection{Overview}
To illustrate and demonstrate the effectiveness of the analysis techniques just described we  apply them to simulated observations of a gravitational wave burst characteristic of the  close, parabolic encounter of two supermassive black holes, such as might occur when the nuclear black holes first find each other following a major merger of two galaxies. We consider four cases: 
\begin{enumerate}
\item A strong signal, for which we can detect the signal, localize the source in the sky, and infer the radiation waveform; 
\item A moderate strength signal, for which we can detect the signal and localize the source, but not accurately infer the waveform;
\item A weak signal, which can be clearly detected but not accurately localized or characterized; and 
\item No signal at all. 
\end{enumerate}
For these examples we use the thirty pulsars in the International Pulsar Timing Array (IPTA) \citep{hobbs:2009:ipt} as described in Table \ref{tbl:ipta}. The measured timing residual for each pulsar is a superposition of white noise with rms timing residual given in Table \ref{tbl:ipta} and red noise normalized to have the same spectral density as the white noise at frequency $0.2\,\mathrm{yr}^{-1}$. Of these thirty pulsars, ten have short-timescale timing residual noise rms less than 0.2$\mu$s, fourteen have short-timescale noise rms between 0.2 and 1$\mu$s and the remaining five have short-timescale noise rms between 1 and 5$\mu$s.\footnote{This is a particular characterization of these pulsars based on communications at the time of this writing from the Parkes Pulsar Timing Array, the European Pulsar Timing Array, and the North American Nanohertz Observatory for Gravitational-waves. It is \emph{not} a definitive characterization.  We are not presenting the data associated with these pulsars but rather using them as an example of a realistic IPTA.}

\begin{deluxetable}{llll}
\tablecolumns{4}
\tablecaption{The International Pulsar Timing Array pulsars, their short-timescale timing noise rms, and the telescopes from which those noise timing residuals were measured.\label{tbl:ipta}}
\tablehead{
\colhead{} & \colhead{Pulsar} & \colhead{RMS ($\mu$s)} & \colhead{Telescope}
}
\startdata
1&J1909-3744&0.054 &GBT \\
2&J1713+0747&0.055 &AO \\
3&J0437-4715&0.060 &Parkes \\
4&J1857+0943&0.066 &AO \\
5&J1939+2134&0.080 &GBT \\
6&J0613-0200&0.110 &GBT \\
7&J1640+2224&0.110 &AO \\
8&J1744-1134&0.130 &GBT \\
9&J1741+1300&0.140 &AO \\
10&J1600-3053&0.190 &GBT \\
11&J1738+0333&0.200 &AO \\
12&J0030+0451&0.300 &AO \\
13&J0711-6830&0.340 &Parkes \\
14&J2317+1439&0.360 &AO \\
15&J2145-0750&0.420 &Parkes \\
16&J1012+5307&0.540 &GBT \\
17&J1022+1001&0.700 &WSRT \\
18&J0218+4232&0.830 &GBT \\
19&J1643-1224&0.880 &Parkes \\
20&J2019+2425&0.910 &AO \\
21&J1024-0719&0.960 &Parkes \\
22&J1455-3330&0.960 &GBT \\
23&1918-0642 &0.960 &GBT \\
24&J1603-7202&0.990 &Parkes \\
25&J2129-5721&0.990 &Parkes \\
26&J1824-2452&1.060 &Parkes \\
27&J1730-2304&1.190 &Parkes \\
28&J1732-5049&1.250 &Parkes \\
29&J1045-4509&1.370 &Parkes \\
30&J2124-3358&2.380 &Parkes \\
\enddata
\end{deluxetable}

The data sets we use for these examples are constructed by 
\begin{enumerate}
\item Calculating the gravitational wave strain associated with the parabolic encounter of two supermassive black holes (see \S\ref{sec:ex1}); 
\item Evaluating the gravitational wave contribution to the pulse arrival time for each pulsar described in Table \ref{tbl:ipta}; 
\item Adding the appropriate noise to the ``gravitational wave'' timing residuals (see \S\ref{sec:nse}); 
\item Removing the best-fit linear trend from the noisy timing residuals.\label{enum:fitOut}
\end{enumerate}
At present, actual pulsar timing residual observations are constructed by fitting actual pulse time-of-arrival data for each pulsar to a timing model characterized by, among other parameters, the pulsar period and period derivative \citep{edwards:2006:tnp}. The final step in the construction of our simulated data --- removing the linear trend --- modifies the data in a manner similar to the ``fitting-out'' procedure that occurs in the construction of actual timing residual data sets. 

To summarize, our simulated data sets model --- in schematic form --- the major features of modern pulsar timing array data sets and the elements that complicate their analysis: white timing noise on short timescales, red timing noise on long timescales, and formation of timing residuals through fitting pulse arrival times to a global timing model. 

\subsection{Construction of simulated data sets}\label{sec:construct}

\subsubsection{Parabolic encounter of two supermassive black holes}\label{sec:ex1}
Following the major merger of two galaxies, each harboring a nuclear supermassive black hole, dynamical friction will drive the nuclear black holes to the nucleus of the merged galaxy. Eventually they will find each other, form a binary, and coalesce.  When they first find each other there may occur a series of close, high-speed encounters, each leading to a burst of radiation, whose duration may be estimated as twice the ratio of the impact parameter to the velocity at periapsis. We adopt this burst as an exemplar for the purpose of demonstrating the effectiveness of the analysis techniques just described gravitational wave burst. At the same time, however, we emphasize that the parabolic encounter gravitational wave model used here is intended as a stand-in for any gravitational wave burst: i.e., the particular model and model parameters adopted here do not correspond to a case we regard as realistic.

We model the parabolic encounter radiation burst via the quadrupole formula applied to the Keplerian parabolic trajectories of the equivalent Newtonian system. 
In the quadrupole approximation the gravitational waves radiated near periapsis are projections of the second time derivative of the systems quadrupole moment: i.e., 
\begin{subequations}
\begin{align}
h_{+} &= \frac{2}{r}\ddot{\mathcal{Q}}^{jk}e^{(+)}_{jk}(\hat{k})\\
h_{\times} &= \frac{2}{r}\ddot{\mathcal{Q}}^{jk}e^{(\times)}_{jk}(\hat{k})
\end{align}
\end{subequations}
where $\hat{k}$ is the unit vector in the direction of wave propagation, we have adopted the Einstein summation convention of summing over repeated indices, and work in units where $G=c=1$. 
For Keplerian parabolic orbits the trajectories (and, correspondingly, the system's quadrupole moment) can be expressed in closed form. Without loss of generality we take the system's motion to be in the $xy$ plane and the periapsis at $y=0$ and $x>0$, in which case
\begin{subequations}
\begin{align}
\ddot{\mathcal{Q}}^{xx} &=
\frac{\mu M}{w_0^3 w_1^4 b}
\left[
-3w_1^3\left(w_1^8 - 6 w_1^6 + 24 w_1^2 -16\right)
+w_0\left(7 w_1^8 - 30w_1^6 + 24 w_1^2 - 16\right)
\right]\\
\ddot{\mathcal{Q}}^{yy} &=
\frac{4\mu M}{w_0^3w_1^2b}
\left[
-3w_1^3\left(w_1^4-4\right)
+ w_0\left(5 w_1^4 - 4\right)
\right]\\
\ddot{\mathcal{Q}}^{xy} &= 
\frac{M \mu}{w_0^3 w_1 b\sqrt{2}}
\left[
w_0\left(-18w_1^4 + 32 w_1^2 + 32\right)
+3 w_1\left(7 w_1^8 -30 w_1^6 + 24 w_1^2 + 16\right)
\right]
\end{align}
where $M$ and $\mu$ are the 
system's total and reduced mass, $b$ is the impact parameters, and 
$\omega_o$ and $\omega_1$ are given by the following:
\begin{align}
w_0 &= \sqrt{8+9\frac{M}{b}\left(\frac{t}{b}\right)^2}\\
w_1 &= \left[3\frac{t}{b}\sqrt{\frac{M}{b}} + w_0\right]^{1/3}.
\end{align}
\end{subequations}
Similarly, the gravitational wave contribution to the timing residual is a projection of the time integral of $\mathbf{h}$ (see Eq.~\ref{eq:tauGW}), which is proportional to the first time derivative of the system's quadrupole moment:
\begin{subequations}
\begin{align}
\dot{\mathcal{Q}}^{xx} &=
\frac{\mu b}{\sqrt{2}w_0 w_1^4}\sqrt{\frac{M}{b}}
\left(w_1^4-4\right)\left(w_1^4-6w_1^2+4\right)\\
\dot{\mathcal{Q}}^{yy} &=
\frac{4\mu b}{w_0w_1^2}\sqrt{\frac{M}{b}}
\left(w_1^4-4\right)\\
\dot{\mathcal{Q}}^{xy} &= 
\frac{b\mu}{\sqrt{2}w_0 w_1^3}\sqrt{\frac{M}{b}}
\left(-3w_1^6 + 8 w_1^4 + 16 w_1^2 - 24\right)
\end{align}
\end{subequations}

\subsubsection{Timing noise}\label{sec:nse}
The millisecond pulsars used in modern pulsar timing arrays typically show white timing noise on short timescales, turning to red noise on timescales of 5--10 years. For the demonstrations here we model the timing noise as the superposition of white noise and red noise, with the red noise contribution normalized to have the same amplitude as the white noise contribution at the frequency $f_{\mathrm{red}}=0.2~\mathrm{yr}^{-1}$. With this normalization the noise power spectrum for each pulsar in our array is completely determined by the short-timescale (white) timing noise rms given in Table \ref{tbl:ipta}.

To compute the red contribution to the timing noise we applying a digital integrator to white noise. To design the integrator we follow \citet{tseng:2006:did}, choosing a single sub-division of the unit delay, a seventh-order FIR filter, and a cascade of three unit delays. The corresponding integrator is given by the transfer function 
\begin{align}
H(z) &= \frac{1}{2^9}\frac{-5+49z^{-1}-245z^{-2}+1225z^{-3}
+1225z^{-4}-245z^{-5}+49z^{-6}-5z^{-7}}{6\left(1-z^{-1}\right)}
\end{align}

\begin{figure}
\plotone{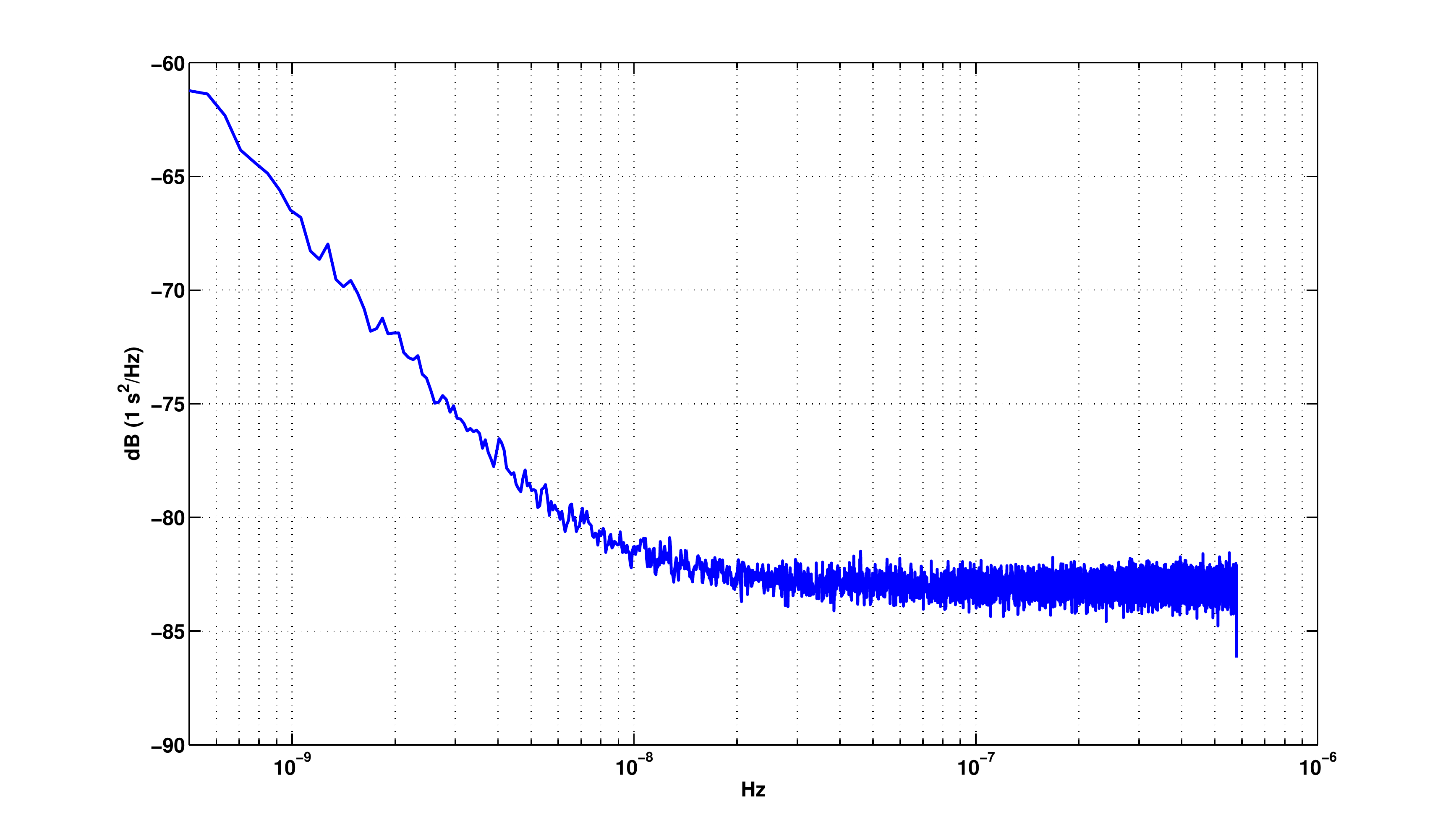}
\caption{Power spectral density of simulated timing noise for PSR J1909-3744. The noise is simulated as the sum of a white noise contribution, which is determining at high frequencies, and a red noise contribution, which is determining at low frequencies. The cross-over frequency is chosen to be 0.2~$\mathrm{yr}^{-1}$, which is characteristic of timing array millisecond pulsars. For more details see \S\ref{sec:nse}.}
\label{fig:psd}
\end{figure}

Figure \ref{fig:psd} shows the characteristics of the power spectral density of the simulated timing noise normalized for PSR J1909-3744. 

\subsection{Analysis of simulated data for strong, moderate, weak, and no signal}\label{sec:analysis}

Our analysis methodology is designed to answer three questions: (1) is a gravitational wave signal present? (2) where is the source? and (3) what is the detailed structure of the waveform? Here we explore the ability of our analysis to answer these questions. For weak signals it may be possible to answer definitively the first of these questions, while being unable to answer the second or third. For stronger signals it may be possible to answer the first question definitively, the second moderately well, and the third not at all. Finally, for the strongest signals all three questions may be answered in detail. We illustrate all three cases in the three following subsections, beginning with a strong signal example and ending with a weak signal example. In each case our ``source'' has waveform characteristic of the parabolic encounter of two $10^9\,\textrm{M}_{\odot}$ black holes, with impact parameter 180\,M (0.02 pc),orbital plane face-on to the Earth line-of-sight, and in the direction of the Virgo cluster (RA~12.5h, dec~12.5$\deg$). Figure \ref{fig:iptaStrongHT} shows, in two panels, the gravitational wave strain incident at Earth (top panel) and the induced timing residuals in a sample of six of the thirty IPTA pulsars when the source is at a distance of 15~Mpc.  We conclude with a subsection exploring how the analysis performs when applied to a data set containing no signal at all.

\subsubsection{Strong signal}\label{sec:ex1-strong}

\begin{figure}
\plotone{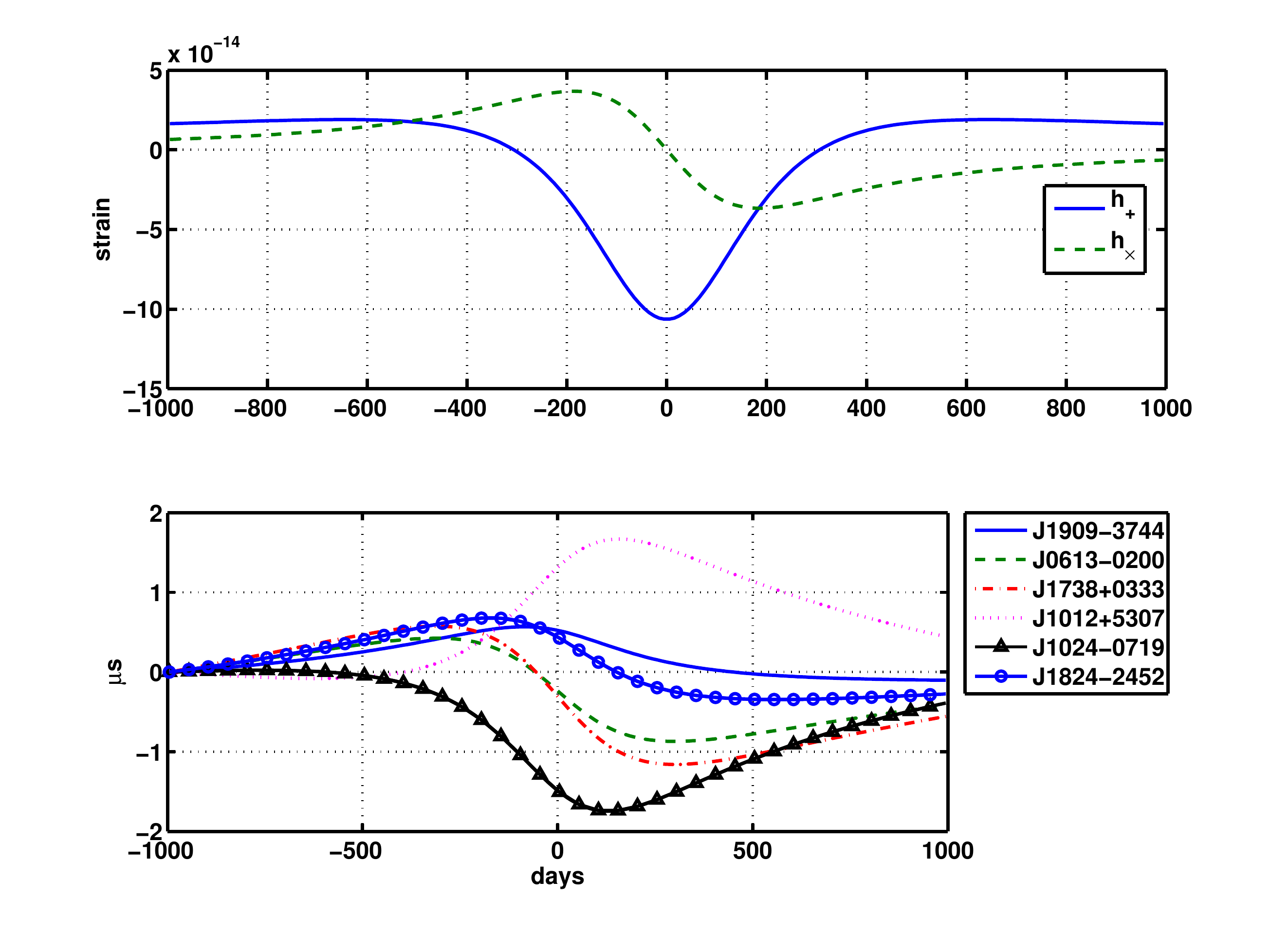}
\caption{The gravitational wave strain incident on Earth and the corresponding timing residuals induced in a sample of International Pulsar Timing Array pulsars. In this example the waves are characteristic of the parabolic encounter of two $10^9~\mathrm{M}_{\odot}$ black holes, impact parameter $180\,\mathrm{M}$ (i.e., 0.02~pc), at a distance of 15~Mpc in the direction of the Virgo Cluster of galaxies (RA 12h5m, Dec 12.5deg). (See discussion of \S\ref{sec:ex1-strong}.) The top panel shows the radiation waveform in the two independent polarization states. The bottom panel shows the timing residuals induced by the waveform in a sample of 6 of the 30 IPTA pulsars.}
\label{fig:iptaStrongHT}
\end{figure}

\begin{figure}
\plotone{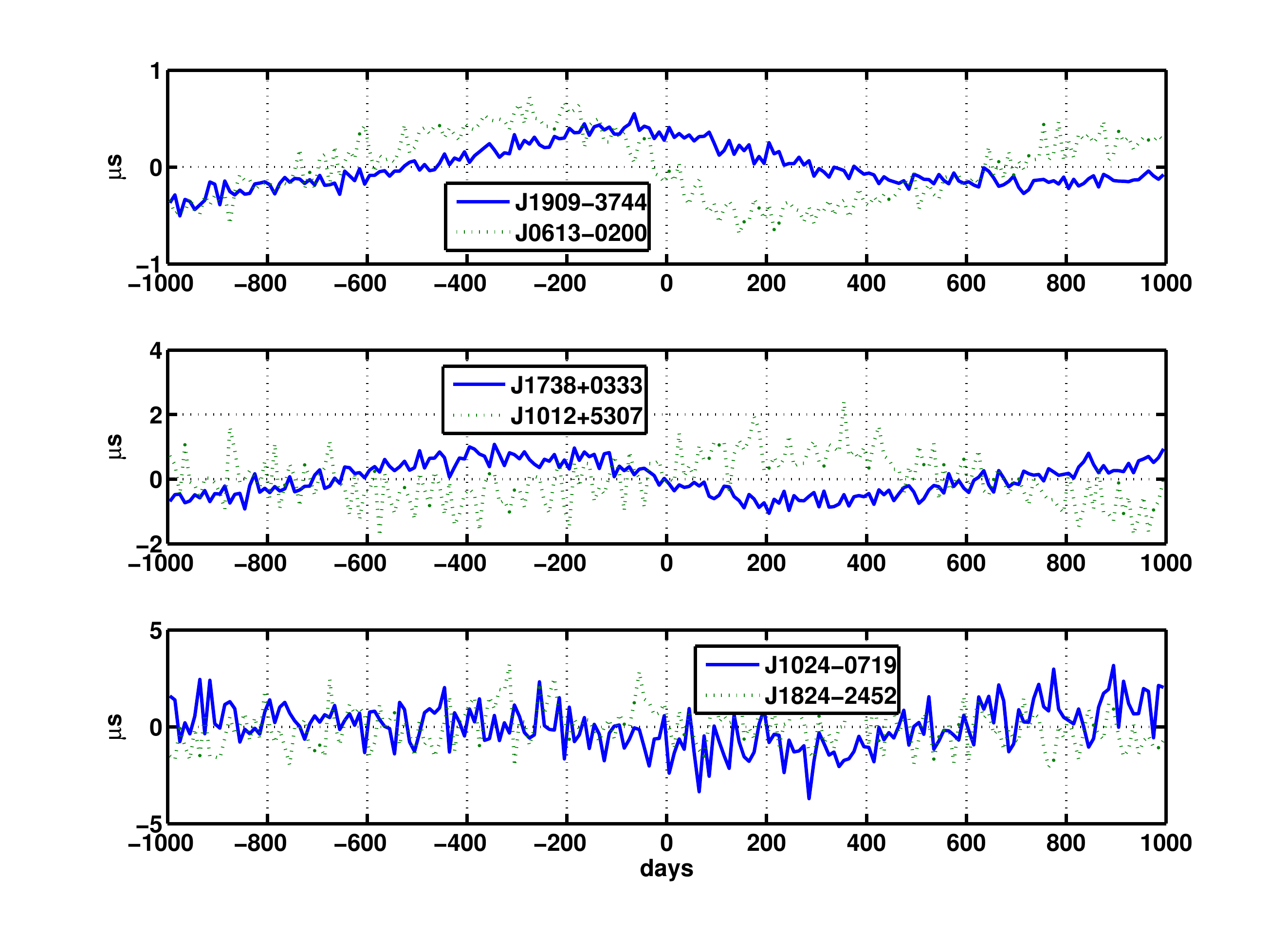}
\caption{The superposition of the gravitational wave induced timing residuals, for the same sample of IPTA pulsars shown in the bottom panel of Fig.~\ref{fig:iptaStrongHT}, with the ``red${}+{}$white'' timing noise (see \S\ref{sec:nse}) characteristic of typical millisecond pulsar timing noise.}
\label{fig:iptaStrongD}
\end{figure}

\begin{figure}
\plotone{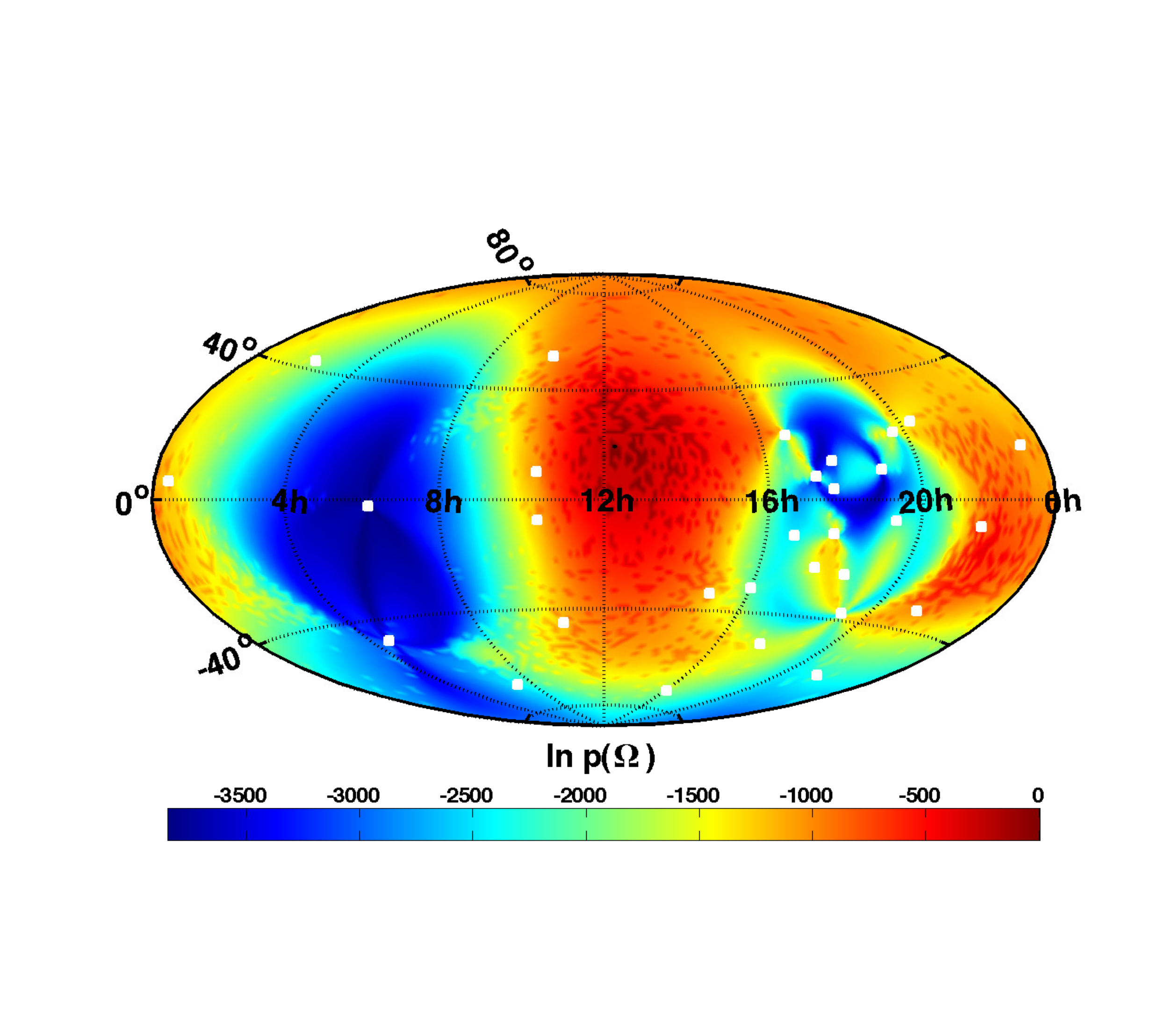}
\caption{Natural log of the inferred probability density that the source of gravitational waves present in the ``strong signal'' simulated IPTA data set described in \S\ref{sec:ex1-strong} is found at location $\Omega$. The smallest 90\% probability contour has an area much less than $1\,\deg^2$ and includes the actual source location. The white squares show the locations of the thirty IPTA pulsars used as detectors.}\label{fig:iptaStrongLocalize}
\end{figure}

\begin{figure}
\plotone{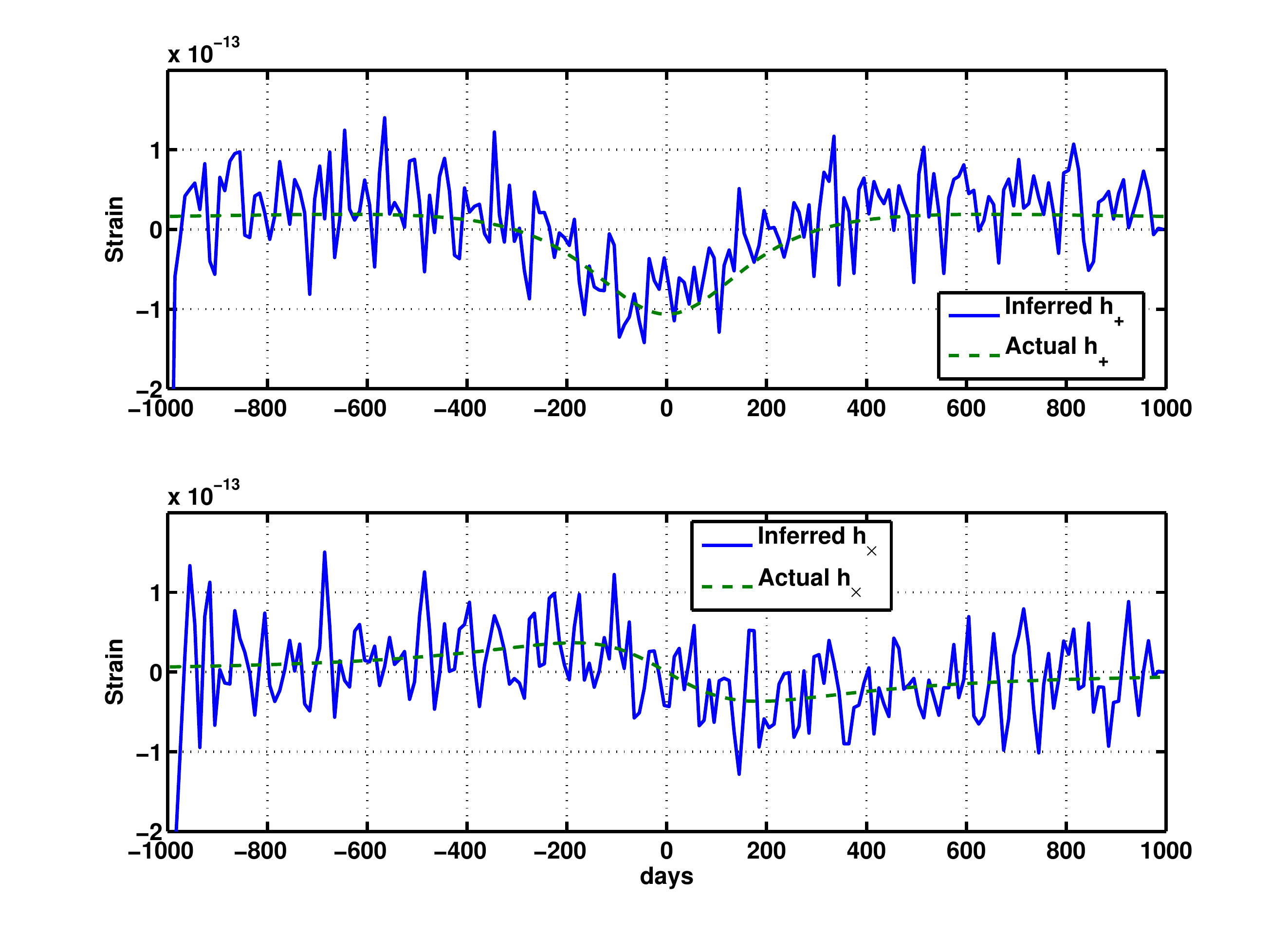}
\caption{The inferred $h_{+}$ and $h_{\times}$ radiation waveforms for the example data set described in \S\ref{sec:ex1-strong}.}
\label{fig:iptaStrongInfer}
\end{figure}

Figures \ref{fig:iptaStrongHT} through \ref{fig:iptaStrongInfer} and the first row of Table \ref{tbl:rslts} summarize the results of applying the methodology described in Section \ref{sec:methodology} to a pulsar timing array dataset including a strong ``flyby'' signal, constructed as described in Section \S\ref{sec:construct}. In this strong signal case the source is placed at a distance of 15~Mpc. The top panel of Figure \ref{fig:iptaStrongHT} shows the strain incident at Earth and the bottom panel the timing residual induced in a sample of six of the thirty IPTA pulsars. Figure \ref{fig:iptaStrongD} shows the same timing residuals, from the same selection of pulsars, embedded in the ``red-plus-white'' timing noise described in \S\ref{sec:nse}. For this strong signal the gravitational wave induced timing residuals are readily apparent in the quietest of the IPTA pulsars (e.g., top two panels of Fig.~\ref{fig:iptaStrongD}), less so in the pulsars with moderate timing noise (middle panels of Fig.~\ref{fig:iptaStrongD}), and much less so in the pulsars with large timing noise (the bottom two panels of Fig.~\ref{fig:iptaStrongD}). 

\begin{deluxetable}{lrrrrr}
\tablecolumns{6}
\tablecaption{Results summary for data analysis applied to four simulated datasets. In all cases the signal corresponds to radiation from a parabolic fly-by of two $10^9\,\mathrm{M}_{\odot}$ black holes propagating from the direction of the Virgo Cluster. In the ``strong'' signal case the source is located at 15~Mpc; in the ``moderate'' signal case the source is at a distance of 100~Mpc; in the ``weak'' signal case the source is at a distance of 260~Mpc; and in the final, ``absent'' signal case the simulated data set consists of timing noise alone. For details see section \S\ref{sec:analysis}.\label{tbl:rslts}}
\tablehead{
\colhead{Signal} & 
\colhead{$\mathrm{ln}\, B(\mathbf{d})$} & 
\colhead{$\Delta\Omega_{90\%}\,(\deg^2)$} &
\colhead{$\rho^2$}& 
\colhead{$\sigma_{+}$} & 
\colhead{$\sigma_{\times}$} 
}
\startdata
Strong&$3.8\times10^3$&$\ll1$&$3.8\times10^{+1}$&$1.1\times10^{-13}$&$1.2\times10^{-13}$\\
Moderate&$6.6\times10^{1}$&$5.8\times10^2$&$8.7\times10^{-1}$&$2.4\times10^{-14}$&$2.4\times10^{-14}$\\
Weak&$2.2\times10^0$&$4.2\times10^3$&$2.1\times10^{-1}$&$2.0\times10^{-14}$&$2.0\times10^{-14}$\\
Absent&$-8.4\times10^0$&$1.2\times10^4$&$7.9\times10^{-2}$&$1.8\times10^{-14}$&$1.9\times10^{-14}$
\enddata
\end{deluxetable}

Applying the analysis described in Section \ref{sec:detect} to this ``strong signal'' dataset instance we find (for our particular instantiation of noise) that the Bayes Factor has a value of $\exp(3.8\times10^3)$, corresponding to overwhelming evidence for the presence of a gravitational wave in this data set. 

Having concluded that a signal is present we use the analysis described in Section \ref{sec:localize} to localize the source. Figure \ref{fig:iptaStrongLocalize} shows the results of this analysis as the natural log of the probability density that the source is in the direction $\Omega$. Also shown is the smallest  contour containing 90\% of the total probability, whose area is much less than $1\,\deg^2$. 

Finally, having detected the source and localized it on the sky, we apply the analysis of Section \ref{sec:infer} to infer the radiation waveform. Figure \ref{fig:iptaStrongInfer} shows the result of this analysis, superposed with the actual radiation waveform. In this example the gravitational wave strain is identified with a power signal-to-noise of 38. 

\subsubsection{Moderate signal}\label{sec:ex1-moderate}

\begin{figure}
\plotone{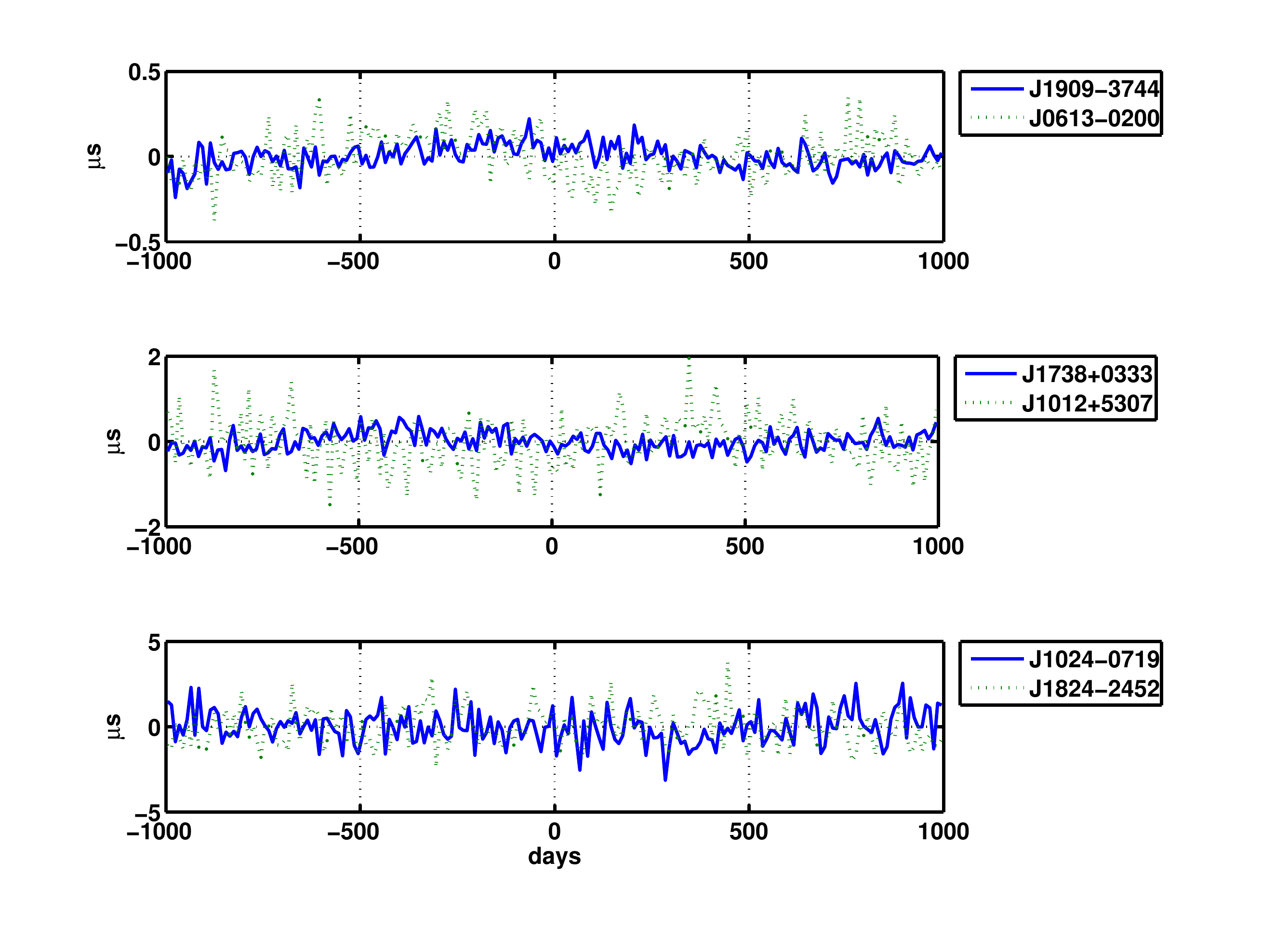}
\caption{The superposition of the gravitational wave induced timing residuals, for the same sample of IPTA pulsars shown in the bottom panel of Fig.~\ref{fig:iptaStrongHT}, with the ``red${}+{}$white'' timing noise (see \S\ref{sec:nse}) characteristic of typical millisecond pulsar timing noise.}
\label{fig:iptaModerateD}
\end{figure}

\begin{figure}
\plotone{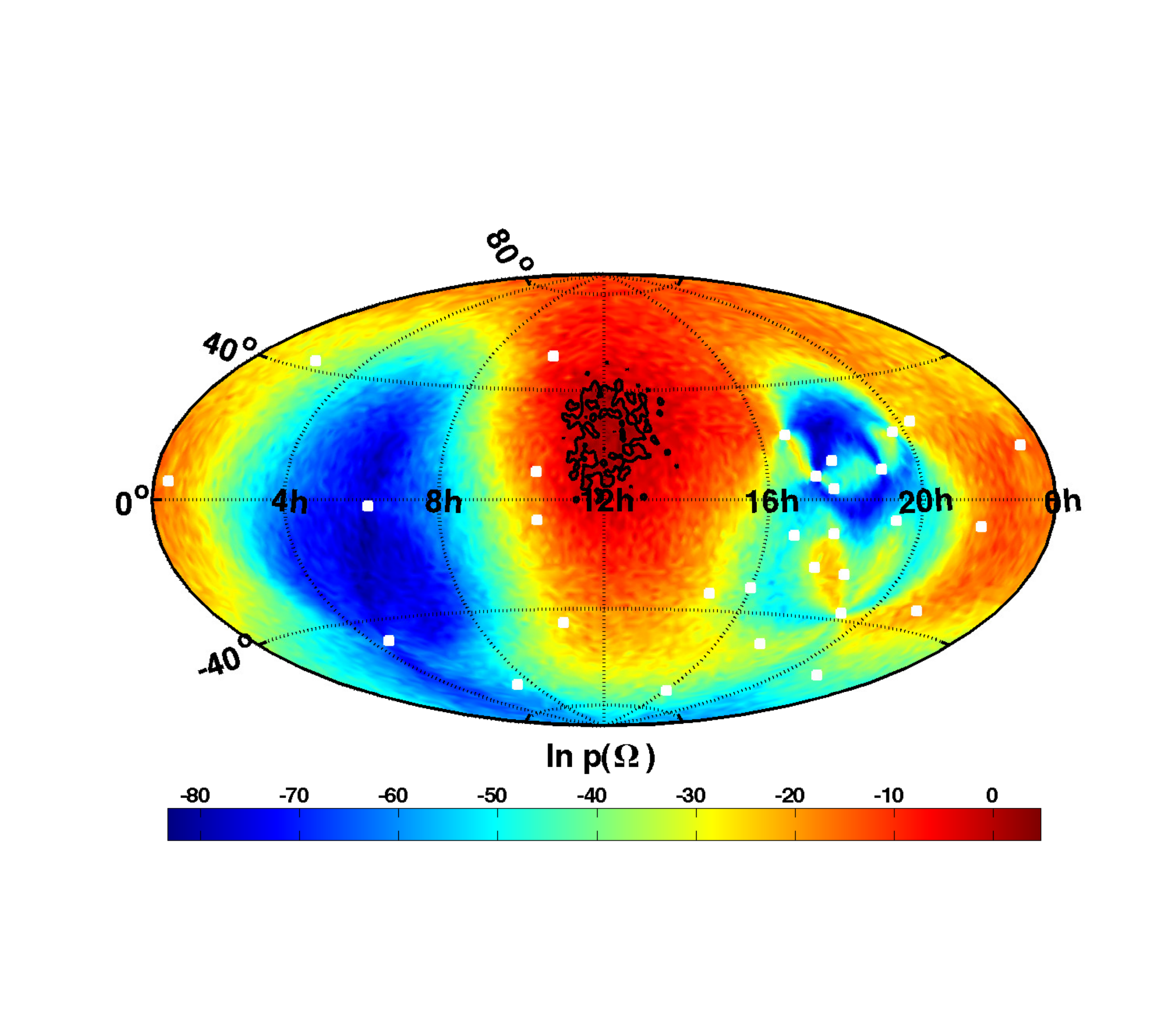}
\caption{Natural log of the inferred probability density that the source of gravitational waves present in the ``moderate signal'' simulated IPTA data set described in \S\ref{sec:ex1-moderate} is found at location $\Omega$. Also shown is the smallest 90\% probability contour, which has an area of $5.8\times10^2~\deg^2$. The white squares show the locations of the thirty IPTA pulsar baselines used as detectors.}\label{fig:iptaModerateLocalize}
\end{figure}

\begin{figure}
\plotone{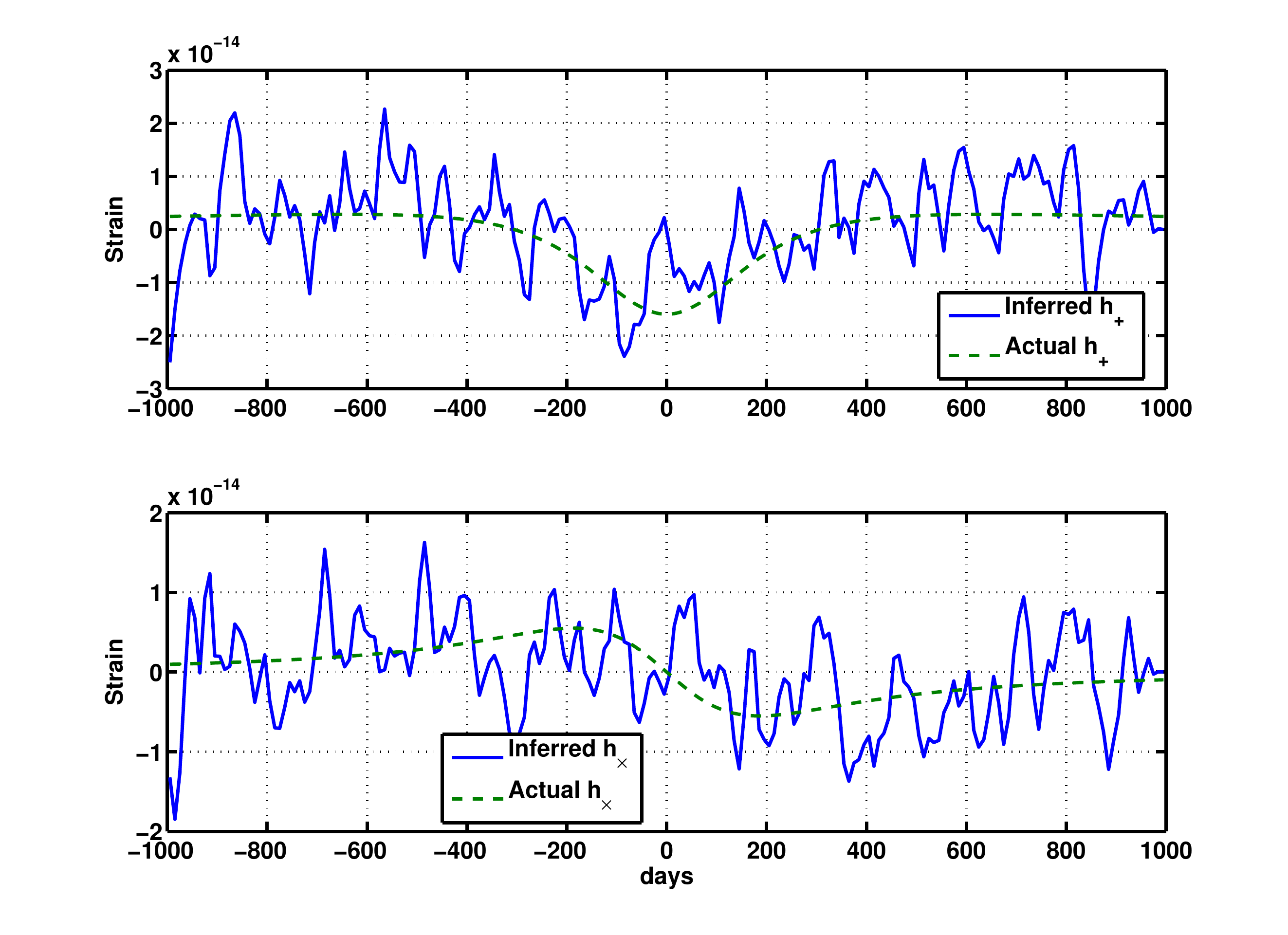}
\caption{The inferred $h_{+}$ and $h_{\times}$ radiation waveforms for the example data set described in \S\ref{sec:ex1-moderate}. While strong enough to be detected the signal is too weak for us to infer its waveform.}
\label{fig:iptaModerateInfer}
\end{figure}

Figures \ref{fig:iptaModerateD} through \ref{fig:iptaModerateInfer} show the results of applying the methodology described in Section \ref{sec:methodology} to a moderate strength ``flyby'' signal observed in the current IPTA. In this case the source is placed at a distance of 100~Mpc in the direction of the Virgo Cluster. Figure \ref{fig:iptaStrongHT}, with the appropriate scaling of the abscissae (i.e., by 15~mpc/100~mpc) shows the gravitational wave strain incident on the IPTA and the corresponding induced timing residuals in a selection of IPTA pulsars. Figure \ref{fig:iptaModerateD} shows the timing residuals, from the same selection of pulsars as in Figure \ref{fig:iptaStrongHT}, embedded in the red-plus-white timing noise described in \S\ref{sec:nse}. For this moderate strength signal the gravitational wave induced timing residuals are apparent in the quietest of the IPTA pulsars (e.g., top two panels of Fig.~\ref{fig:iptaModerateD}), but not apparent in the residuals with of the other pulsars.

Applying the analysis described in Section \ref{sec:detect} to this data set we find that the Bayes Factor has a value of $\exp(66.)$, again corresponding to overwhelming evidence for the presence of a gravitational wave signal. 

Having concluded that a signal is present we attempt to localize the source using the analysis described in Section \ref{sec:localize}. Figure \ref{fig:iptaModerateLocalize} shows the results of our localization analysis as the log of the probability density that the source is in the direction $\Omega$. Contours enclosing the smallest area containing $90\%$ of the total probability are also shown. These contours, which encloses an area of $5.8\times10^2\,\deg^2$, correctly include the actual source location.

Finally, having detected the source and localized it on the sky, we apply the analysis of Section \ref{sec:infer} to infer the radiation waveform. Figure \ref{fig:iptaModerateInfer} shows the result of this analysis, made assuming we know the actual source location, superposed with the actual radiation waveform. In this example the power signal-to-noise ratio is 0.87, which confirms the impression given by the figure that this inference is not significant. 

This moderate signal amplitude case shows clearly a regime where the gravitational wave burst is strong enough to be unambiguously detected and the general direction to the source  clearly identified (even if not so precisely that an optical counterpart may be sought), but not strong enough to characterize the burst waveform. 

\subsubsection{Weak signal}\label{sec:ex1-weak}
\begin{figure}
\plotone{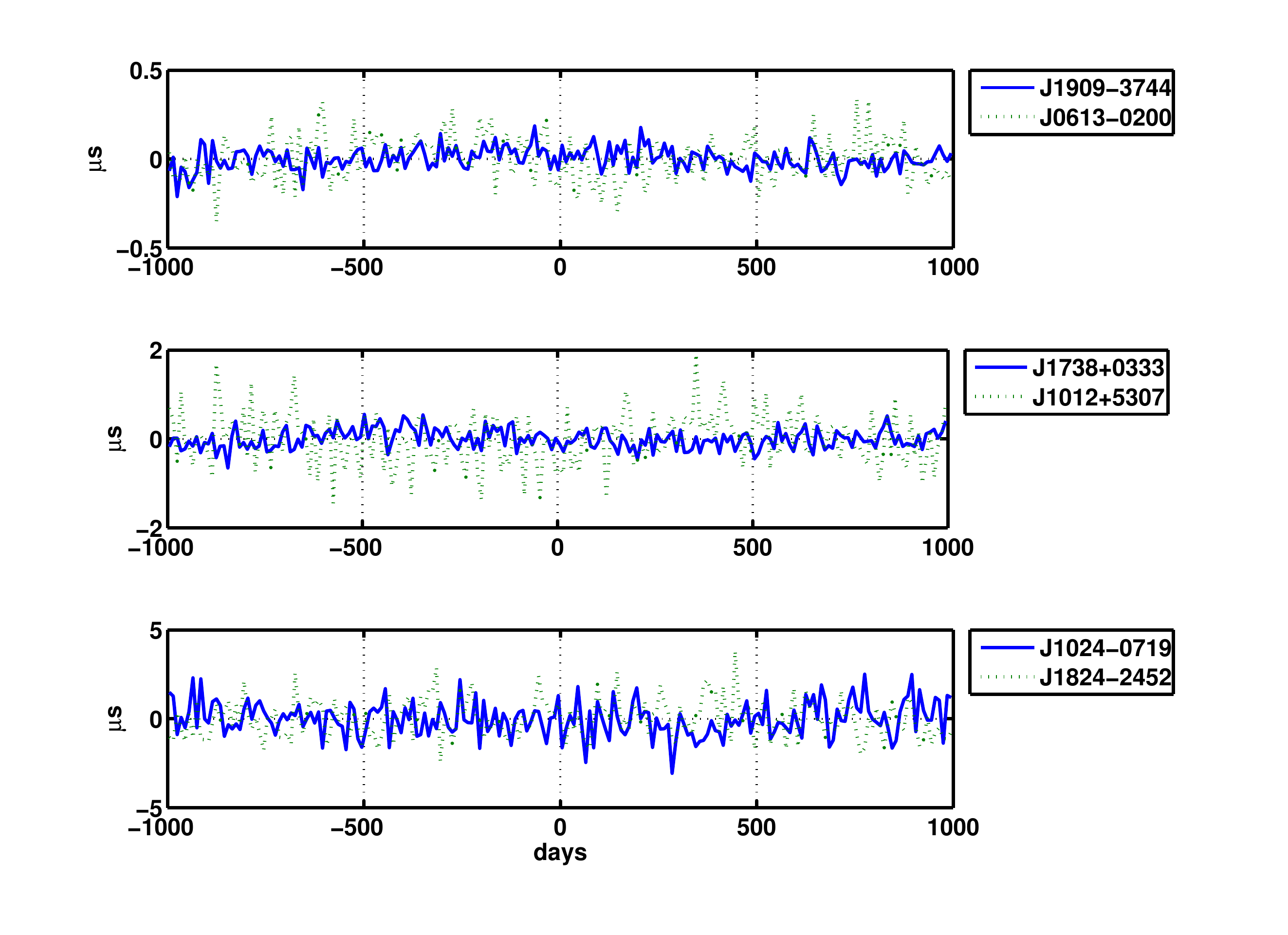}
\caption{The superposition of the gravitational wave induced timing residuals, for the same sample of IPTA pulsars shown in the bottom panel of Fig.~\ref{fig:iptaStrongHT}, with the ``red${}+{}$white'' timing noise (see \S\ref{sec:nse}) characteristic of typical millisecond pulsar timing noise.}
\label{fig:iptaWeakD}
\end{figure}

\begin{figure}
\plotone{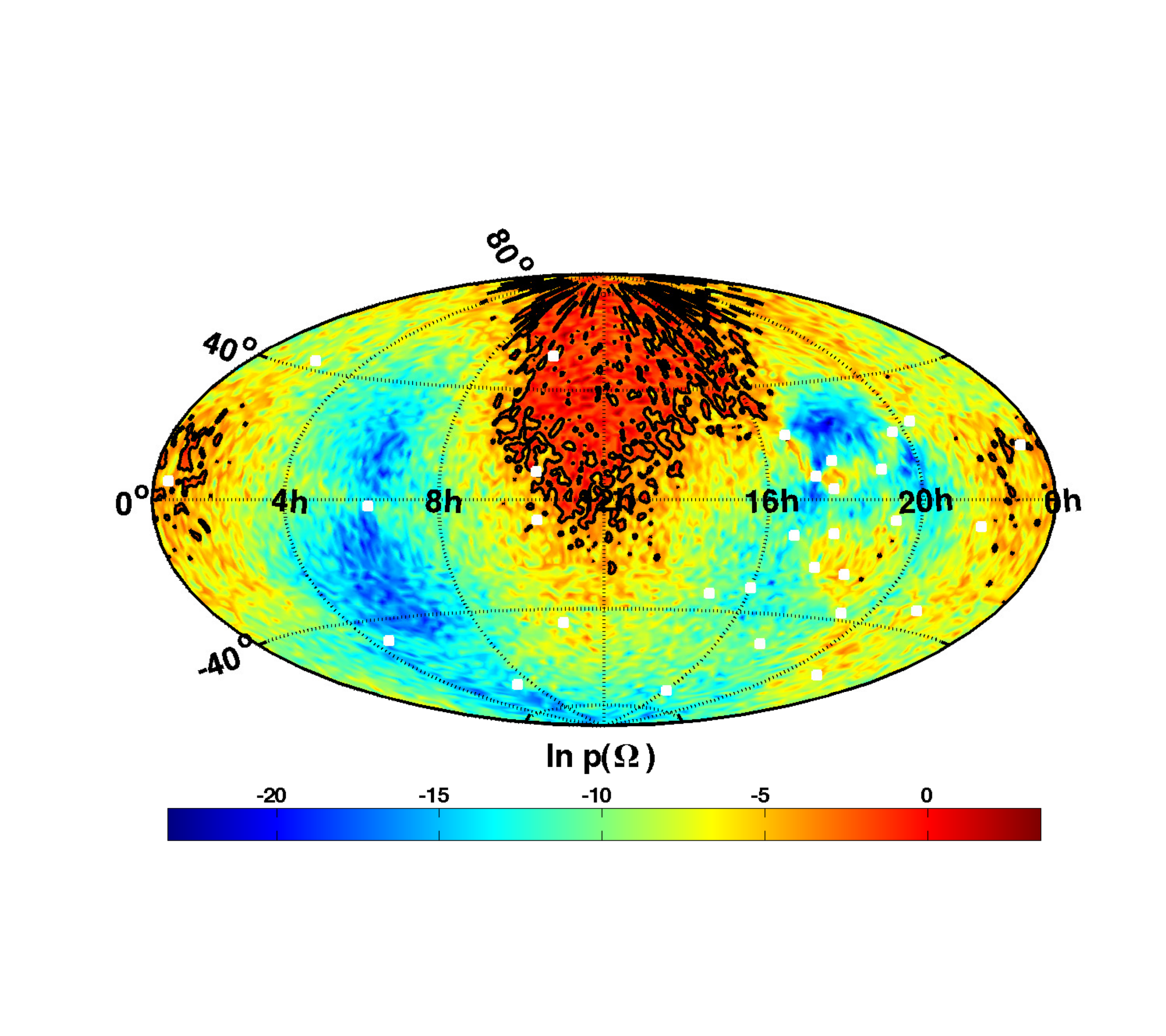}
\caption{Natural log of the inferred probability density that the source of gravitational waves present in the ``weak signal'' simulated IPTA data set described in \S\ref{sec:ex1-weak} is found at location $\Omega$. While strong enough to be detected, the signal is too weak to be reliably localized: the 90\% probability contour has an area of $4.2\times^3\deg^2$. The white squares show the locations of the thirty IPTA pulsar baselines used as detectors.}\label{fig:iptaWeakLocalize}
\end{figure}

\begin{figure}
\plotone{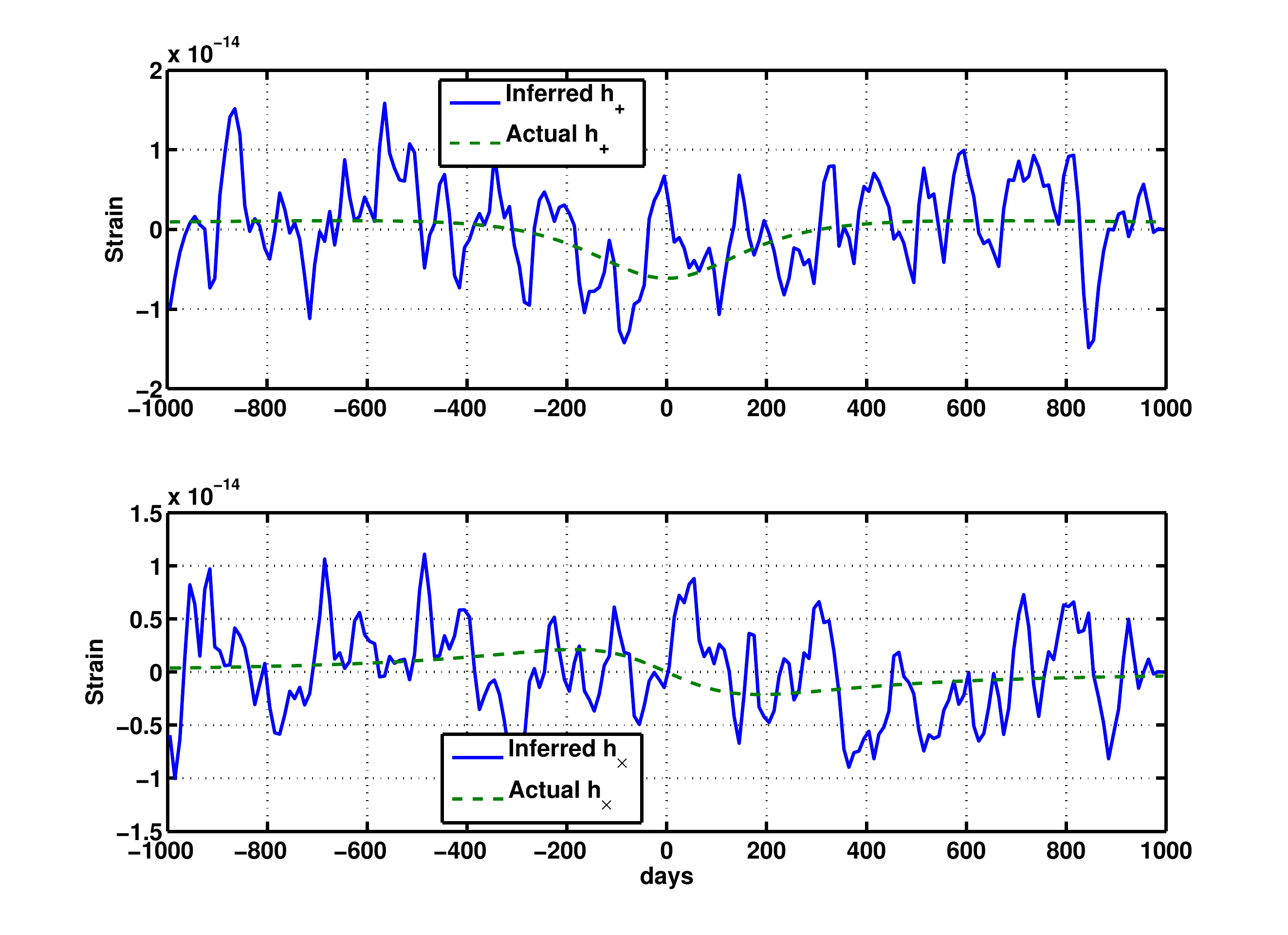}
\caption{The inferred $h_{+}$ and $h_{\times}$ radiation waveforms for the example data set described in \S\ref{sec:ex1-weak}. While strong enough to be detected, the signal is too weak to for us to infer its waveform.}
\label{fig:iptaWeakInfer}
\end{figure}

Finally, we consider a dataset that includes a signal at the edge of detectability: i.e., a data set where the Bayes Factor corresponds to 9:1 odds of a signal being present, with results shown in Figures \ref{fig:iptaWeakD} through \ref{fig:iptaWeakInfer}. In this case our binary source is placed at a distance of $2.6\times10^2$~Mpc. Figure \ref{fig:iptaStrongHT}, with the appropriate scaling of the abscissae (i.e., by 15~kpc/261~mpc) shows the gravitational wave strain incident on the IPTA and the corresponding induced timing residuals in a selection of IPTA pulsars. 
Figure \ref{fig:iptaWeakD} shows the timing residuals from the same selection of pulsars as in Figure \ref{fig:iptaStrongHT}, embedded in white timing noise with rms given in Table \ref{tbl:ipta}. For this weak signal the gravitational wave induced timing residuals are not readily apparent even in the quietest pulsars (e.g., top panel of Fig.~\ref{fig:iptaWeakD}). 

Applying the analysis described in Section \ref{sec:detect} to this data set we find that the Bayes Factor has a value of $\exp(2.2)=9$. At this level our prejudice regarding the likelihood of gravitational wave bursts passing through our timing array plays a critical role in deciding whether the overall odds --- i..e, the product of the Bayes Factor with the ``expectation odds'' --- are in favor of detection or not. Supposing that they are we next attempt to localize the source using the analysis described in Section \ref{sec:localize}. Figure \ref{fig:iptaWeakLocalize} shows the results of our localization analysis as the log of the probability density that the source is in the direction $\Omega$. In this case, the $90\%$ contour encloses an area of $4.2\times10^3\,\deg^2$ scattered about the sky: i.e., the wave, while strong enough to be detected, is not strong enough to be localized. 

Finally, and for completeness, we apply the analysis of Section \ref{sec:infer} to infer the radiation waveform. Figure \ref{fig:iptaWeakInfer} shows the result of this analysis, made assuming that we know the actual source location on the sky, superposed with the actual radiation waveform. In this example the power signal-to-noise ratio is $0.21$. 

\subsubsection{No signal}\label{sec:none}
\begin{figure}
\plotone{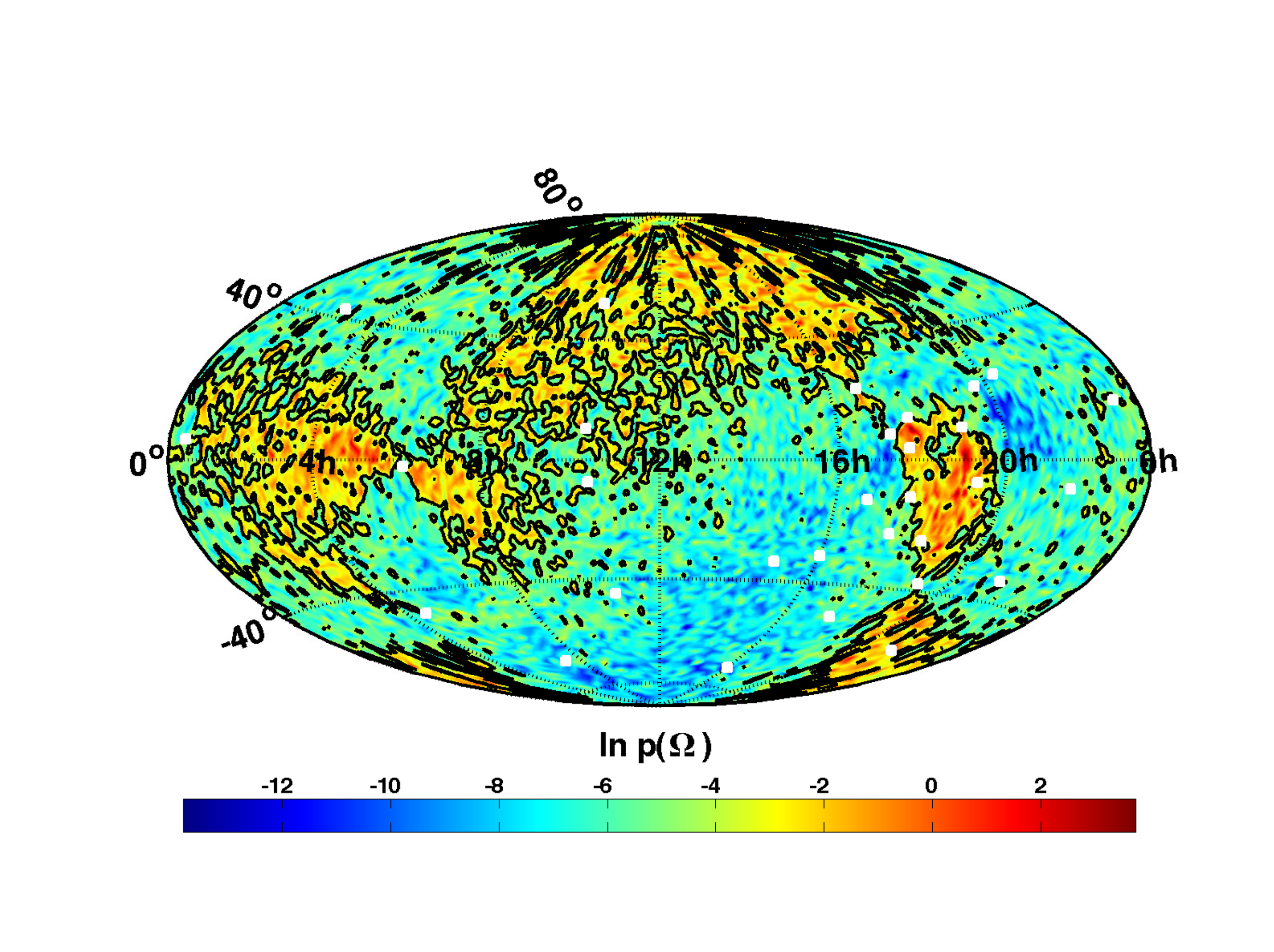}
\caption{Natural log of the inferred probability density that a source of gravitational waves is present in the direction $\Omega$ for a data set consisting \emph{only} of noise.  The 90\% probability contour has an area of $1.2\times^4\deg^2$. The white squares show the locations of the thirty IPTA pulsar baselines used as detectors. See \S\ref{sec:none} for more details.}\label{fig:iptaNone}
\end{figure}

\begin{figure}
\plotone{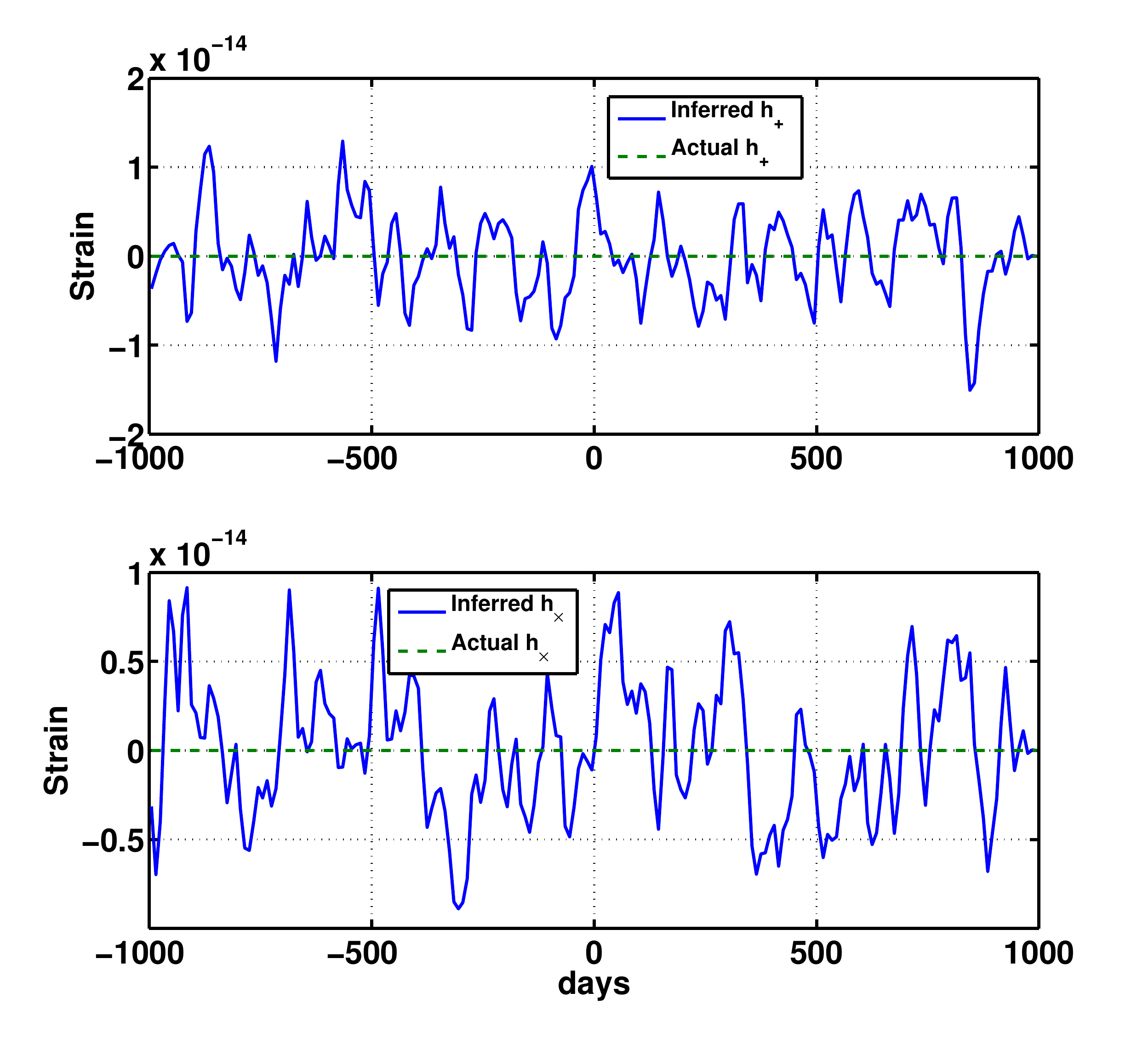}
\caption{The inferred $h_{+}$ and $h_{\times}$ radiation waveforms for a data set consisting only of timing noise. The inferred waveform corresponds to a signal-to-noise of $7.8\times10^{-2}$ See \S\ref{sec:none} for further discussion.}
\label{fig:waveNone}
\end{figure}

Finally, we apply our analysis to a data set consisting of noise only. In this particular instance the Bayes Factor is $\exp(-8.4)=2.2\times10^{-4}$: i.e., overwhelming evidence for the \emph{absence} of a gravitational wave burst. For completeness, under the assumption that a single source is present we show in Figure \ref{fig:iptaNone} the inferred probability density for the source location on the sky. In this case the 90\% confidence interval has an area of $1.2\times10^4\,\deg^2$: i.e., approximately 1/3 of the sky. Lastly, under the assumption that there is a source in the direction of the Virgo Cluster we attempt to infer a waveform from this data set. Figure \ref{fig:waveNone} shows the results of this analysis, which correspond to a power signal-to-noise ratio of $7.8\times10^{-2}$. We conclude that the analysis described here is fully capable of identifying data sets that contain no evidence of a gravitational wave signal. 

\section{Conclusions}
\label{sec:conclusions}

In the history of astronomy few (if any) new observational windows have been as eagerly anticipated as the gravitational wave window, whose opening will provide us with a novel and direct view of astronomical phenomena that can now be inferred at best dimly and indirectly.  Making sense of what we see through this new window requires analysis tools and techniques adapted to the unique nature of our new ``telescopes'' and the sources they enable us to study. Here we have described an intra-related suite of analysis techniques for gravitational wave astronomy designed to address quantitatively three specific questions: 
\begin{enumerate}
\item What are the ``odds'' that a gravitational wave detector data set includes the signal from a gravitational wave burst?
\item Assuming that a gravitational wave burst is present in a data set, what is the probability that the wave is propagating in direction $\hat{k}$? 
\item Assuming the presence of a burst propagating in direction $\hat{k}$, what is the probability that the wave at Earth is characterized by the functions $h_{+}(u)$ and $h_{\times}(u)$, $u = t-\hat{k}\cdot\vec{x}$, representing the $+$ and $\times$ polarization state waveforms?
\end{enumerate}

We address these questions in the specific context of gravitational wave detection using pulsar timing array data. Until recently, analyses for gravitational wave detection using timing data from an array of pulsars has focused on stationary sources: e.g., a stochastic gravitational wave signal or the signal from a binary system. By addressing burst sources we also add to the very recent literature examining how pulsar timing data can be used to detect gravitational wave bursts \citep{haasteren:2009:gma} such as might arise a close fly-by or collision of two supermassive black holes or from a cosmic string cusp \citep{binetruy:2009:gwb,key:2009:cgw}. 

To demonstrate the efficacy of our analysis we applied it to four synthetic timing residual data sets  representative of observations using the International Pulsar Timing Array \citep[IPTA]{hobbs:2009:ipt}. Each data set included simulated timing noise, constructed to be characteristic of actual IPTA timing noise. Three of the datasets included the timing residual signature of a gravitational wave burst characteristic of a parabolic fly-by of two supermassive black holes; the fourth did not. The three ``signal-present'' cases varied only by the gravitational wave signal amplitude. In the case of the strongest signal the burst was unambiguously detected, localized to much better than a $\deg^2$, and the waveform in the individual polarization states recovered. In the moderate signal amplitude case the signal was, again, unambiguously detected and the general direction to the source clearly determined; however, the signal amplitude was too low to infer the waveform characteristics. In the third case the signal was strong enough to be detected but too weak to be characterized or to allow the source to be localized. Finally, in analyzing noise alone the calculated odds were, as they should be, unambiguously against the presence of a gravitational wave burst.  

At present, pulsar timing array data sets are constructed by fitting a timing mode to the time of arrival (TOA) data for each pulsar. This timing model includes, in parameterized form, all the non-gravitational-wave contributions that affect the pulse arrival times. The residual differences between the timing model predictions and the actual arrival times are then analyzed for the signature of a passing gravitational wave. This procedure has the disadvantage of being incapable of identifying any gravitational wave whose effect on the arrival time of individual pulsars is degenerate with any of the non-gravitational-wave effects that are part of the timing model. Our analysis may be extended to infer, in addition to the gravitational radiation waveform, the other timing model parameters. This extended analysis may be applied to TOA data directly, avoiding entirely the problem of ``fitting-out'' gravitational wave contributions whose character is similar to other timing model contributions. We intend to investigate this extension in future work. 

While our presentation and discussion have focused on pulsar timing array observations the analysis methodology that we describe applies equally well and without modification to gravitational wave data taken from ground-based detector networks like the LIGO-Virgo network \citep{accadia:2010:sap,riles:2010:pgw}, or for the analysis of LISA \citep{jennrich:2009:trl,merkowitz:2009:cls} data. 

\acknowledgments

We thank Lynn Baker for valuable conversations, and Martin Hendry and Graham Woan for comments on an early version of the manuscript. This work was supported by National Science Foundation grants AST-0748580 (AL) and PHY 06-53462 (LSF).

\end{document}